\documentclass[notitlepage,showkeys,nofootinbib,11pt,tightenlines,superscriptaddress
]{revtex4-2}
\usepackage{placeins}
\usepackage{amsmath,comment,subfigure}
\usepackage{placeins}
\usepackage{enumitem}
\usepackage{tikz}

\usepackage{siunitx,soul,,xcolor}
\setstcolor{red}
\usepackage{graphicx}
\usepackage{color,makecell,adjustbox,multirow,colortbl}
\usepackage{hhline}
\usepackage{setspace}

\usepackage{dcolumn}
\usepackage{bm}
\makeatletter
\let\ps@titlepage\ps@plain
\makeatother
\usepackage[utf8]{inputenc}
\usepackage[T1]{fontenc}
\usepackage{mathptmx,epstopdf} 
\usepackage[colorlinks=true,linkcolor=blue]{hyperref}
\hypersetup{
colorlinks=true,    
linkcolor=blue,       
citecolor=blue,       
filecolor=blue,        
urlcolor=blue,        
linktoc=page          
}

\usepackage{tabularray}
\usepackage{paralist}
\usepackage{pythonhighlight}
\usepackage{color,makecell,adjustbox,multirow}
\usepackage{booktabs} 
\usepackage{hhline}
\begin{document}
\onecolumngrid
\title{Machine Learning-Enhanced Design of Lead-Free Halide Perovskite Materials Using Density Functional Theory}

\author{Upendra Kumar }
\altaffiliation{These authors contributed equally to this work}
\affiliation{Division of Carbon Neutrality and Digitalization, Korea Institute of Ceramic Engineering and Technology (KICET), Jinju 52851, South Korea.}
\author{Hyeon Woo Kim}
\altaffiliation{These authors contributed equally to this work}
\affiliation{Division of Carbon Neutrality and Digitalization, Korea Institute of Ceramic Engineering and Technology (KICET), Jinju 52851, South Korea.}
\affiliation{Department of Materials Science and Engineering, Hanyang University, Seoul 04763, South Korea.}
\author{Gyanendra Kumar Maurya}
\affiliation{Department of Physics, GLA University, Mathura 281406, India.}
\author{Bincy Babu Raj}
\affiliation{Division of Carbon Neutrality and Digitalization, Korea Institute of Ceramic Engineering and Technology (KICET), Jinju 52851, South Korea.}
\affiliation{School of Material Science and Engineering, Pusan National University, 2 Busandaehak-ro 63beon-gil, Geumjeong-gu, Busan 46241, South Korea.}
\author{Sobhit Singh}
\affiliation{ Department of Mechanical Engineering at the University of Rochester, New York 14611, United States.}
\affiliation{Materials Science Program, University of Rochester, Rochester, New York 14627, USA}
\author{Ajay Kumar Kushwaha}
\affiliation{Department of Metallurgy Engineering and Materials Science, Indian Institute of Technology Indore, Khandwa Road, Simrol, Indore 453552, India.}
\author{Sung Beom Cho}
\email{csb@ajou.ac.kr}
\affiliation{Department of Energy Systems Research, Ajou University, Suwon 16499, South Korea.}
\author{Hyunseok Ko}
\email{hko@kicet.re.kr}
\affiliation{Division of Carbon Neutrality and Digitalization, Korea Institute of Ceramic Engineering and Technology (KICET), Jinju 52851, South Korea.}


\begin{abstract}
\noindent 
The investigation of emerging non-toxic perovskite materials has been undertaken to advance the fabrication of environmentally sustainable lead-free perovskite solar cells. This study introduces a machine learning methodology aimed at predicting innovative halide perovskite materials that hold promise for use in photovoltaic applications. The seven newly predicted materials are as follows: CsMnCl$_4$, Rb$_3$Mn$_2$Cl$_9$, Rb$_4$MnCl$_6$, Rb$_3$MnCl$_5$, RbMn$_2$Cl$_7$, RbMn$_4$Cl$_9$, and CsIn$_2$Cl$_7$. The predicted compounds are first screened using a machine learning approach, and their validity is subsequently verified through density functional theory calculations.
CsMnCl$_4$ is notable among them, displaying a bandgap of 1.37 eV, falling within the Shockley-Queisser limit, making it suitable for photovoltaic applications. Through the integration of machine learning and density functional theory, this study presents a  methodology that is more effective and thorough for the discovery and design of materials.
\end{abstract}
\maketitle
\vspace{-0.8cm}

\section{Introduction}
Halide perovskites have emerged as promising candidates for revolutionizing the landscape of solar cell technology \cite{yin2015halide}. Characterized by their unique crystalline structure and exceptional optoelectronic properties, halide perovskites offer unparalleled potential for efficient and cost-effective solar energy conversion. Their tunable bandgap \cite{unger2017roadmap}, high absorption coefficients \cite{lei2021metal}, and long carrier diffusion lengths \cite{shrestha2022long} make them highly attractive for photovoltaic applications. Moreover, their facile synthesis methods \cite{jathar2021facile} and compatibility with flexible substrates \cite{dai2020scalable} open avenues for scalable and versatile solar cell designs. Beyond their application in solar cells, halide perovskite materials hold immense promise for a wide range of electronic devices \cite{mathur2021organolead}. However, concerns over the toxicity of lead-based halide perovskites have prompted significant research efforts towards developing lead-free alternatives. \par Lead-free perovskite halide materials \cite{ning2019structural} represent a promising avenue in the quest for sustainable and environmentally friendly solar energy solutions. These materials offer potential for high-performance solar cells \cite{chen2020high} while mitigating environmental and health risks. Research efforts have focused on exploring a diverse range of lead-free perovskite compositions, including tin, bismuth, and other metal halides \cite{jin2020critical}, which exhibit encouraging optoelectronic properties and demonstrate potential for efficient photovoltaic applications \cite{hoefler2017progress}. By addressing the need for lead-free halide designs, researchers aim to unlock the full potential of halide perovskite materials while ensuring environmental sustainability and human health safety.
\par  Since only a few elements around lead in the periodic table have been considered, there are still many unexplored compositional regions in lead-free perovskites. This is due to the difficulty of experimental trial and error. Therefore, there is a need for techniques that can accelerate material design, such as machine learning (ML). Research efforts have focused on synthesizing lead-free perovskite halides with enhanced stability, efficiency, and device performance \cite{ma2020lead}. Despite challenges in achieving commercial viability and scalability, ongoing advancements in material engineering and device fabrication techniques are paving the way for the integration of lead-free perovskite halides into next-generation electronic devices \cite{pecunia2020lead}. 
\par  Recent efforts have been made to identify lead-free halide perovskite materials through data mining from the Materials Project \cite{jain2013commentary}.	The Debye temperature and thermal stability also have implications for thermal management in electronic devices \cite{cui2020emerging}. By selecting materials with appropriate Debye temperatures, engineers can design electronic components that operate efficiently within specified temperature ranges without experiencing degradation or failure. This aspect is particularly crucial in the development of high-performance computing systems and advanced electronics. Hence, we compute the Debye temperature for the dataset comprising halide perovskites. Within this investigation, we have uncovered overlooked halide perovskite  and predicted novel compounds suitable for photovoltaic purposes.
ML techniques \cite{CGCNN} are utilized for the prediction of vast number number of halide perovskite materials. To confirm this prediction of the forecasted  halide perovskite materials by ML, density functional theory (DFT) is utilized. This integrated method of combining ML with DFT offers a greater opportunity for streamlined and thorough method for the discovery and design of materials, aiding in the recognition of halide perovskite materials possessing sought-after electronic and optical properties suitable for photovoltaic applications.
\section{Methods}
\noindent \textbf{ML Technique}\\
The  generalized Crystal Graph Convolutional Neural Networks (CGCNN) framework, developed by T. Xie and J. C. Grossman \cite{CGCNN}, is tailored to effectively represent periodic crystal systems. They \cite{CGCNN} utilized eight distinct properties in their research: \textbf{(i)} crystallographic information, \textbf{(ii)} formation energy, \textbf{(iii)} absolute energy, \textbf{(iv)} bandgap, \textbf{(v)} Fermi energy, \textbf{(vi)} bulk moduli, \textbf{(vii)} shear moduli, and \textbf{(viii)} Poisson ratio. In their CGCNN model, crystallographic information was used as a feature, while the remaining seven properties served as targets. In the CGCNN model, structures are represented as graphs where the nodes correspond to atoms and the edges represent the bonds between them. The features used for these nodes are derived from elemental properties, which include atomic number, electronegativity, ionization energy, atomic radius, and other fundamental attributes of the elements. These features provide critical information that captures the chemical and physical characteristics of each atom, enabling the model to learn complex relationships and interactions within the crystal structure for accurate property prediction. The CGCNN  model is trained via Crystallographic Information Files (CIF).  A CIF is a standard text file format used to describe the crystallographic structure of a material. It contains detailed information about the atomic coordinates, symmetry operations, unit cell parameters, and other essential data needed to define the crystal structure. CIFs are widely used in materials science and crystallography to facilitate data exchange and enable accurate structure determination and analysis.
Remarkably, CGCNN's interpretability is exhibited through the extraction of energy contributions specific to sites within perovskite structures, offering insights into how local chemical environments influence materials properties. \\ 
\noindent \textbf{Dataset and Features}\\
We have effectively obtained an extensive dataset from the Materials Project \cite{jain2013commentary}. Hence, we conducted extensive data mining for halogen ternary compounds (perovskite halide), specifically involving Br, Cl, and I, leveraging the vast Materials Project database \cite{jain2013commentary}.   In the collated dataframe includes crucial physical quantities such as \textbf{(i)} material identification, \textbf{(ii)} pretty-formula representation, \textbf{(iii)} bandgap values, \textbf{(iv)} convex hull information, \textbf{(v)} space-group symbols, \textbf{(vi)} Inorganic Crystal Structure Database (ICSD) identifiers, \textbf{(vii)} structural details, \textbf{(viii)} elasticity characteristics, \textbf{(xi)} Crystallographic Information File (CIF) data, and \textbf{(x)} Debye temperature values ($\Theta_{\mathrm{D}}$). The $\Theta_{\mathrm{D}}$ and bandgap  dataset encompasses a total of 456 and 2608 rows, respectively. Our objective was to swiftly predict key material properties i.e. $\Theta_{\mathrm{D}}$ and bandgap for  of lead-free halide perovskites.
Both datasets consist of 10 columns, rendering it a valuable asset for conducting varied research and analyses in the realm of materials science. The CGCNN model's features stem from the crystal graph, depicted through crystallographic information files, offering a thorough portrayal of the structural intricacies of the crystal system.  Approximately 70\% of the data  is designated for training. Another 15\% is reserved for testing, and an equal 15\% is allocated for validation. The CGCNN model undergoes a total of 30 epochs.
\par
\begin{figure}[hbt!]
\begin{center}
\includegraphics[width=0.60\textwidth,height=0.40\textwidth]{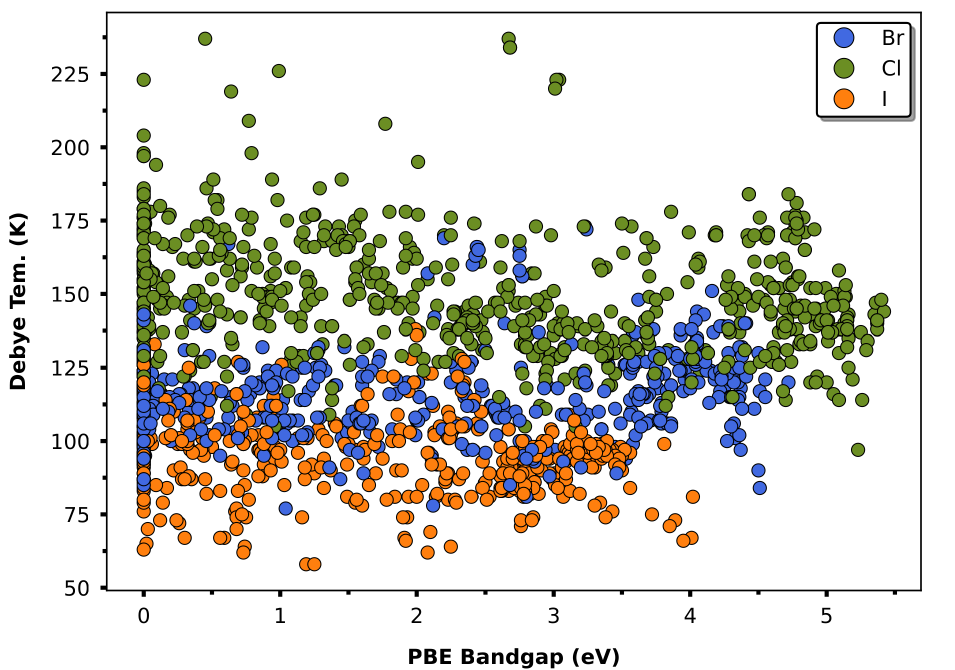}
\caption{\small The plot illustrates the correlation between Debye temperature and bandgap for ternary perovskite halides based on Br, Cl, and I.}
\label{Debye_BandGap}
\end{center}
\end{figure}

 For screening materials, we used three criteria: \textbf{(i)} the material should be a \mbox{perovskite} ternary halide, \textbf{(ii)} it must be lead-free, and \textbf{(iii)} its bandgap should fall within the Shockley-Queisser limit. A high $\Theta_{\mathrm{D}}$ indicates strong atomic bonds, leading to superior thermal conductivity and stability at elevated temperatures \cite{denault2014consequences}. A larger bandgap provides better electronic insulation, making the material suitable for high-power and high-frequency applications. The data of $\Theta_{\mathrm{D}}$ is very limited in the Materials Project database \cite{jain2013commentary}. Hence, we conducted extensive data mining for halogen ternary compounds (perovskite halide), specifically involving Br, Cl, and I, leveraging the vast Materials Project database \cite{jain2013commentary}. Our objective was to swiftly predict key material properties i.e. $\Theta_{\mathrm{D}}$ and bandgap of lead-free halide perovskites. To achieve this, we employed the CGCNN ML model \cite{CGCNN}, which has demonstrated remarkable efficacy in predicting material properties based on crystal structures. After performing data mining, it is evident from Fig.(\ref{Debye_BandGap}) that Cl-based materials in the lead-free halide perovskite family exhibit both a large bandgap and high $\Theta_{\mathrm{D}}$. Therefore, we focused on predicting new structures within Cl-based materials in this lead-free halide perovskite family.

\par \noindent \textbf{DFT Method}\\ 
All the DFT calculations reported in this work are conducted using the Vienna Ab Initio Simulation Package (VASP) employing the projector augmented wave method (PAW) \cite{kresse1996efficiency,kresse1996efficient}. The exchange and correlation energies are calculated using the generalized gradient approximation parameterized by Perdew-Burke-Ernzerhof (PBE) \cite{PBE}. A plane wave cut-off energy of 400 eV was utilized for all DFT calculations. The force convergence criterion for relaxing inner-atomic coordinates was set to \mbox{$10^{-4}$ eV$/\textup{~\AA}$}, and the energy convergence criterion for self-consistent DFT calculations was established at 10$^{-6}$ eV. 
\begin{figure}[hbt!]
\centering
\includegraphics[width=0.8\linewidth]{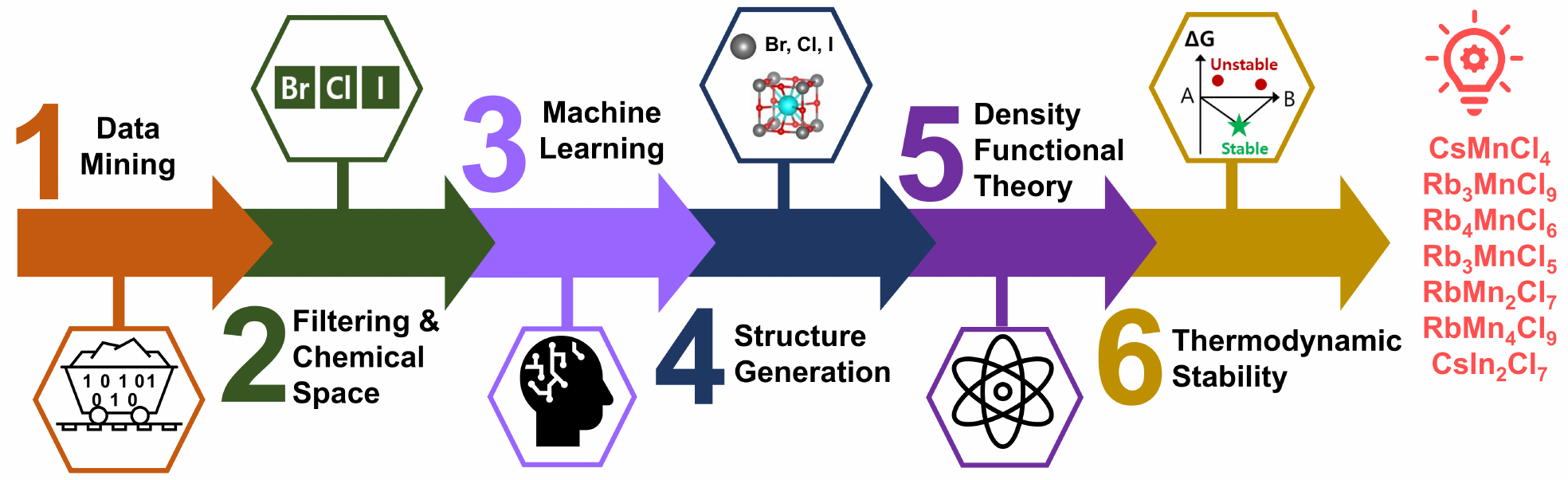}
\caption{\small Flowchart schematic outlining our methodology for discovering new materials in this work.}
\label{workflow1}
\end{figure}	
\par A detailed flowchart illustrating our workflow is presented in Fig.(\ref{workflow1}). Our method starts initiates with data mining \textbf{(1)}, involving the thorough examination of extensive databases to find relevant information concerning known materials and their photovoltaic characteristics. Subsequently, the exploration and filtration of chemical space \textbf{(2)} are conducted to pinpoint possible choices based on specific criteria. The incorporation of ML technique \textbf{(3)} enhance predictive capabilities by extrapolating patterns from the accumulated data. Subsequently, the creation of crystal structures \textbf{(4)} is performed to propose novel material configurations. The utilization of density functional theory \textbf{(5)} is crucial in evaluating the electronic and optical properties of these structures, ensuring their suitability for photovoltaic application. Ultimately, thermodynamic stability analysis \textbf{(6)} is utilized to confirm the viability of the proposed materials, providing a comprehensive and rigorous strategy for anticipating new photovoltaic materials with improved precision and performance. 
\section{Result and Discussion}
\par Our findings unveiled intriguing trends among the studied halogen family members. Specifically, chloride-based materials exhibited a notably large bandgap and high Debye temperature compared to the bromide and iodide counterparts. This discovery provides valuable insights into the potential suitability of chloride-based halide perovskites for certain applications where these properties are advantageous. 
The predictive accuracy of the CGCNN model for bandgap and Debye temperature was quantified, demonstrating an impressive 83\% accuracy for bandgap prediction and 90\% accuracy for Debye temperature prediction. These results underscore the robustness and reliability of the CGCNN model in rapidly screening and predicting material properties.
\begin{figure}[hbt!]
\centering
\includegraphics[width=0.9\linewidth]{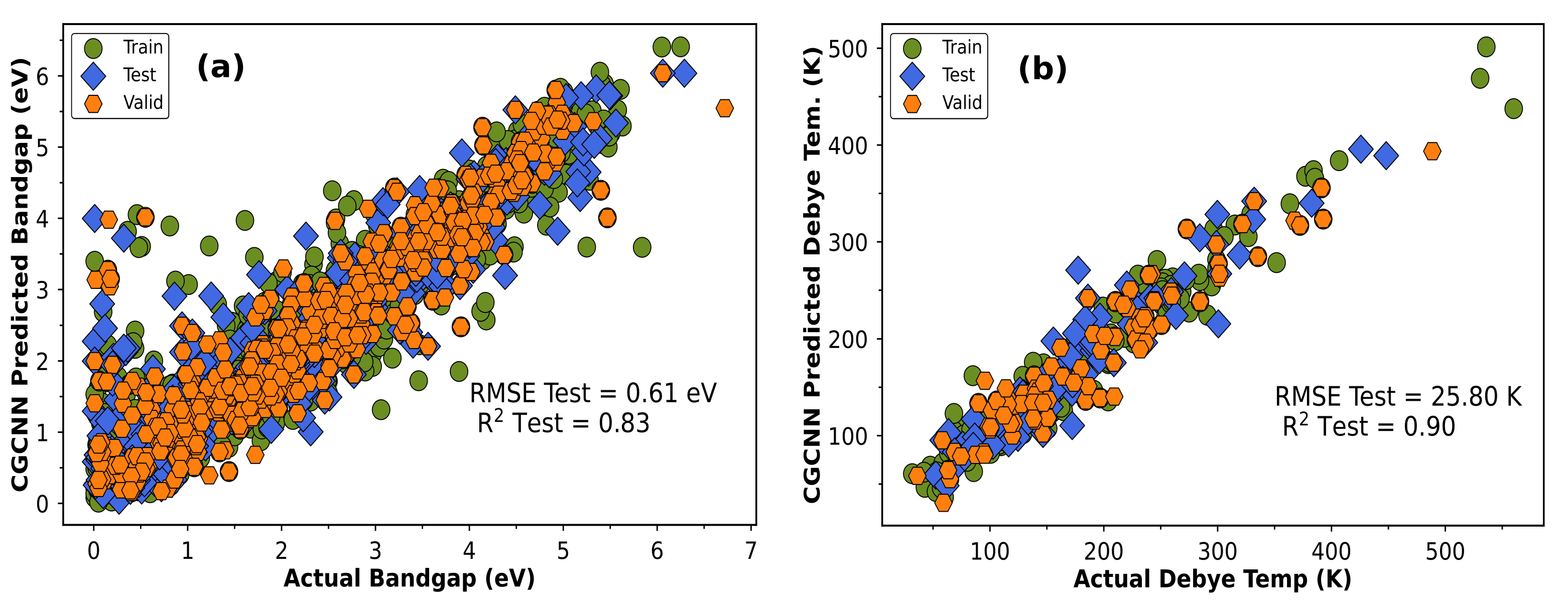}
\caption{ \small The comparison plot illustrates the disparity between actual values (DFT-PBE) and CGCNN-predicted values for both \textbf{(a)} bandgap and \textbf{(b)} Debye temperature.}
\label{Debye_BandGap1}
\end{figure}
\par Hautier et al.\cite{hautier2011data} introduced a novel probabilistic model designed to evaluate the feasibility of ionic species substituting for one another while preserving the crystal structure. This pioneering model leverages an extensive experimental database of crystal structures for its training. Inspired by techniques from the field of machine translation, this model incorporates substitution knowledge into a specialized probability function. This function can be employed to quantitatively gauge the probability that a particular substitution will result in the formation of another stable compound maintaining the same crystal structure.
Recently, H. Kim et al. adopted a same approach, as described by Hautier et al., to predict the emergence of a novel halide perovskite, specifically Cs$_3$LuCl$_6$ \cite{kim2023high}. This underscores the continued utility and effectiveness of the model in predicting new compounds with desired crystal structures in the realm of materials science. We provide the code used for the structure predictor, given in {\color{blue}{ Supplementary Section(I)}}. A similar approach has been also applied in Ref.\cite{kumar2024designing} new Pr based material i.e. Pr$_3$AlO$_6$ , Pr$_4$Al$_2$O$_9$, Pr$_3$ScO$_6$ and Pr$_3$Sc$_5$O$_{12}$ exhibit a large bandgap and high Debye temperature. 

\par The Cs-Mn-Cl, Rb-Mn-Cl, and Cs-In-Cl families bandgap  dataset encompasses a total number of 2608 value. The data mining is performed by using materials project \cite{jain2013commentary}. In this study, we apply the equivalent ML algorithm to predict new structures in the Cs-Mn-Cl family, Rb-Mn-Cl family, and Cs-In-Cl family. The ML model identifies seven new compounds, namely \textbf{(i)} CsMnCl$_4$, \textbf{(ii)} Rb$_3$Mn$_2$Cl$_9$, \textbf{(iii)} Rb$_4$MnCl$_6$, \textbf{(iv)} Rb$_3$MnCl$_5$, \textbf{(v)} RbMn$_2$Cl$_7$, \textbf{(vi)} RbMn$_4$Cl$_9$, and \textbf{(vii)} CsIn$_2$Cl$_7$, all of which have zero energy above the convex hull, calculated by DFT. Detailed information about these materials is provided in Table(\ref{BandGapDebye}) and \textcolor{blue}{Supplementary Section (II)}. 
\par
\begin{figure}[h]
\centering
\includegraphics[width=0.9\linewidth]{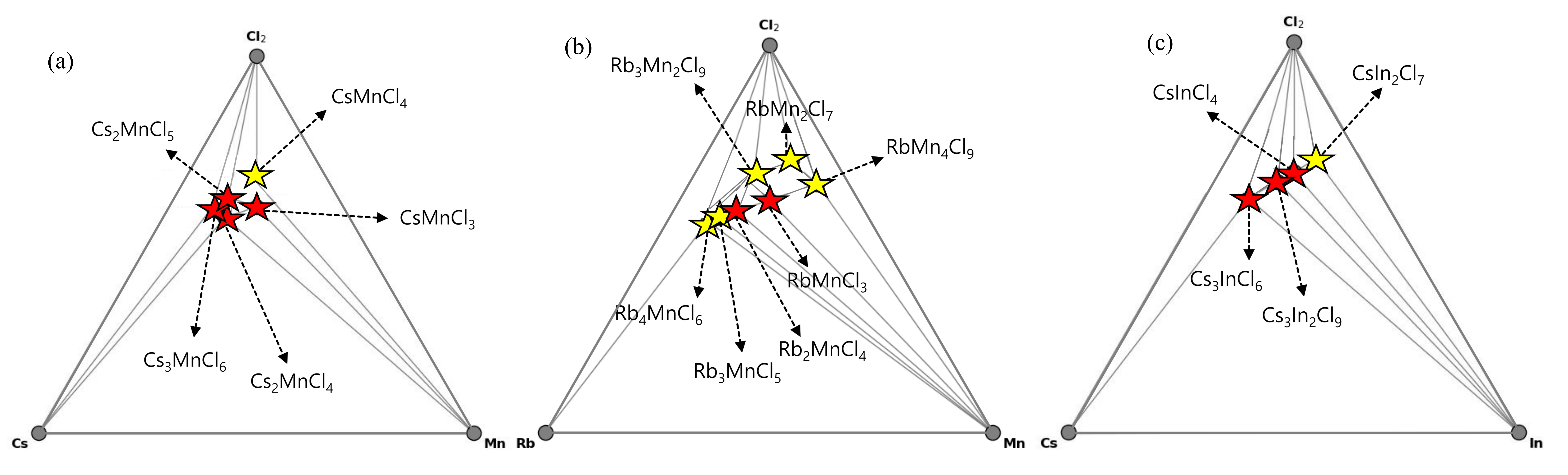}
\caption{\small The convex hull plot is shown for \textbf{(a)} the Cs-Mn-Cl family, \textbf{(b)} the Rb-Mn-Cl family, and \textbf{(c)} the Cs-In-Cl family. Known compounds from the materials-project database are marked with red stars \cite{jain2013commentary}, while newly predicted compounds are represented by magenta stars.}
\label{convex_hull_plot}
\end{figure}

 Calculating the formation energy and constructing the associated convex hull is a fundamental step in determining the energetic stability of a compound \cite{peterson2021materials}. From a thermodynamic perspective, the convex hull pertains to the Gibbs free energy of the compounds at absolute zero. Our computations demonstrate the energetic stability of the newly predicted compounds, as validated by the convex hull plot presented in Fig.(\ref{convex_hull_plot}). Additional details, encompassing E$_{\mathrm{hull}}$ and parent atom information, can be found in \textcolor{blue}{Supplementary Section (III)}.
\subsection{Bandgap and Density of States} 
\begin{figure}[hbt!]
\centering	
\includegraphics[width=0.97\linewidth]{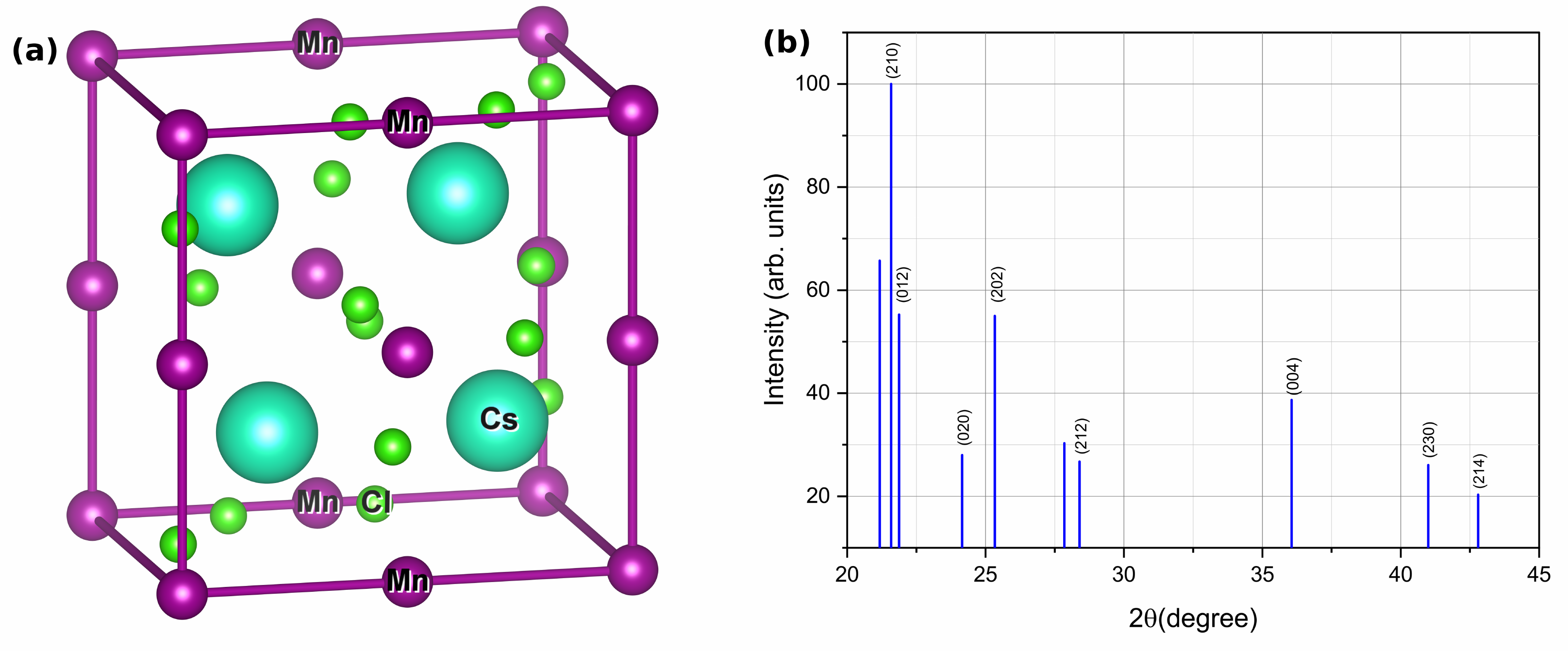}
\caption{\small  CsMnCl$_4$ \textbf{(a)}  crystal structure (\textsl{P2$_1$/c}), \textbf{(b)} simulated-XRD data generated utilizing Pymatgen’s XRD module.}
\label{XRD_Ball_Stick}
\end{figure}

The electronic bandgap of the newly predicted compounds is computed utilizing both PBE-DFT \cite{PBE} and CGCNN \cite{CGCNN} methods. The computed bandgap values for all candidates are documented in \mbox{Table (\ref{BandGapDebye})}. The Debye temperature for the newly predicted compounds was determined using CGCNN \cite{CGCNN} and validated against DFT calculations, as detailed in Table (\ref{BandGapDebye}). All of the predicted compounds exhibit a substantial bandgap and elevated Debye temperature. Our focus is on lead-free materials with bandgaps falling within the Shockley-Queisser limit \cite{shockley2018detailed}, typically ranging from approximately 1.1 eV to 1.6 eV. Among the seven predicted compounds, only CsMnCl$_4$ possesses a bandgap of 1.37 eV, placing it within the Shockley-Queisser limit. Consequently, we delve into a detailed discussion of the material properties of CsMnCl$_4$. The structure detail and simulated x-ray diffraction spectrum  \cite{ling2022solving} of CsMnCl$_4$, shown in Fig.(\ref{XRD_Ball_Stick}).
\par In the context of the Shockley-Queisser limit \cite{shockley2018detailed}, the indirect bandgap imposes limitations on the radiative recombination process, leading to lower photogenerated current and efficiency. Researchers are actively exploring innovative approaches, such as tandem structures and advanced materials, to overcome the limitations posed by the indirect bandgap and push the efficiency of silicon solar cells closer to the Shockley-Queisser limit. This pursuit of efficiency improvements in indirect bandgap semiconductors is crucial for the continued advancement of solar cell technologies and the broader goal of sustainable energy production. Hence, the predicted material CsMnCl$_4$ consists a potential suitability for photovoltic applications.

\begin{table}[hbt!]
\centering
\scalebox{0.90}{\begin{tabular}{|c|c|c|c|c|c|c|c|c|c|}
\hline
&&&&&&\\
{\quad \quad } &  \thead{\textbf{\quad Compound \quad}\\ {}}  &  \thead{\textbf{Bandgap}\\ {\textbf{\quad DFT-PBE (eV) \quad}}}  & \thead{\textbf{Bandgap}\\ {\textbf{CGCNN (eV)}}} &  \thead{\textbf{\quad Bandgap \quad} \\ \textbf{Type}} & \thead{\textbf{\quad Debye Temp. \quad}\\ \textbf{\quad DFT-PBE (K) \quad}} & \thead{\textbf{Debye Temp.}\\\textbf{CGCNN (K)}} \\
&&&&&&\\
\hline
&&&&&&\\
1  &     CsMnCl$_4$ &      1.37&1.54 &   Indirect&147.23&125.57\\
&&&&&&\\
\hline
&&&&&&\\
2  &   Rb$_3$Mn$_2$Cl$_9$ &      0.00&0.23 &   Indirect & 140.86&142.71\\
&&&&&&\\
\hline 
&&&&&&\\
3  &    Rb$_4$MnCl$_6$ &  2.61&2.34 &    Direct &178.65&173.35\\
&&&&&&\\
\hline 
&&&&&&\\
4  &    Rb$_3$MnCl$_5$ &      2.84&2.44 &   Indirect &158.96&160.19\\
&&&&&&\\
\hline 
&&&&&&\\
5  &    RbMn$_2$Cl$_7$  &      0.79&0.95&   Indirect &142.22&145.55\\
&&&&&&\\
\hline 
&&&&&&\\
6  &    RbMn$_4$Cl$_9$  &      1.89&1.15 &   Indirect &163.49&154.51\\
&&&&&&\\
\hline
&&&&&&\\
7  &     CsIn$_2$Cl$_7$  &      3.01&3.18 &   Indirect &139.25&121.85\\
&&&&&&\\
\hline
\end{tabular}}
\caption{\small The newly predicted materials underwent analysis to determine their bandgap and Debye temperature. The CGCNN model was trained using data generated from DFT-PBE calculations.
}
\label{BandGapDebye}
\end{table}
\par
\begin{figure}[hbt!]
\centering	
\includegraphics[width=1.0\linewidth]{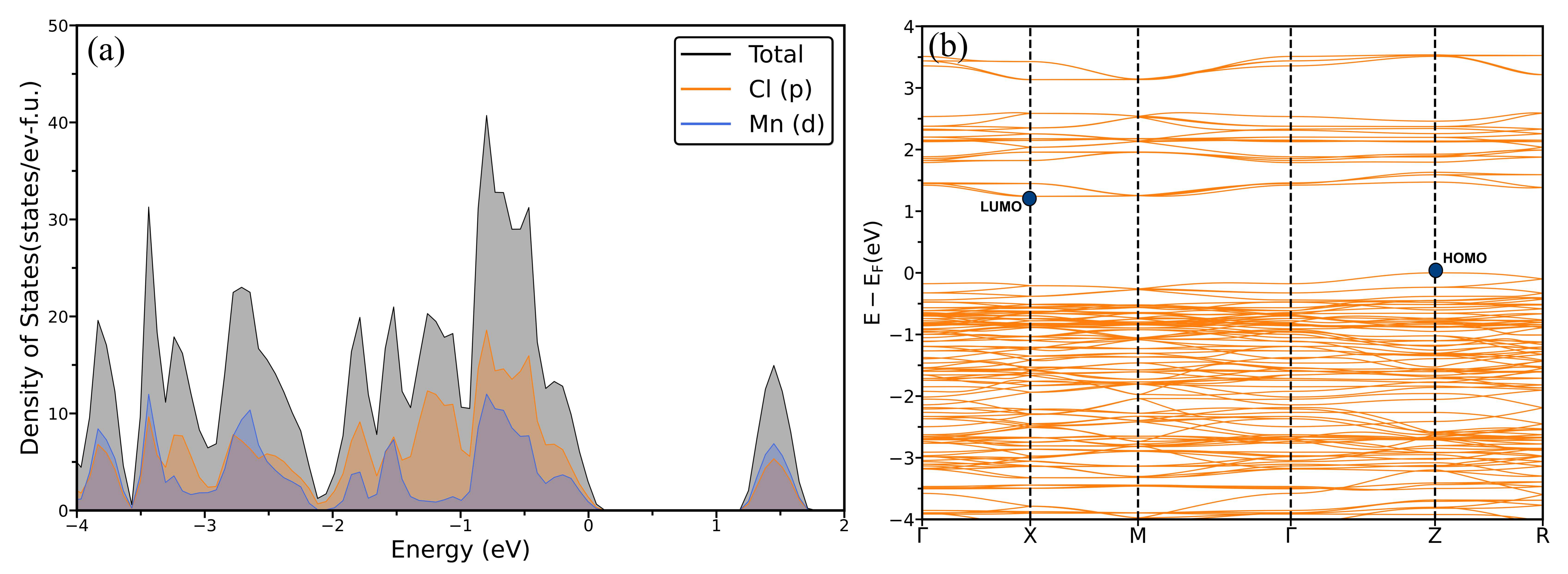}
\caption{\small The PBE calculated  \textbf{(a)} DOS, \textbf{(b)} band structure, the HOMO and LUMO, showing it is an indirect bandgap. }
\label{dosbandcollage}
\end{figure}

  To gain further insights into the electronic structure of the newly predicted compound CsMnCl$_4$, we computed both the total electronic density of states (DOS) and the atomic-orbitals resolved electronic DOS, as depicted in Fig.\ref{dosbandcollage}\textbf{(a)}. In the anticipated compound CsMnCl$_4$, the predominant contributions arise from the $p$ orbitals of chlorine and the $d$ orbitals of manganese. The peak near -1 eV can be attributed to the hybridization of all constituent elements. In the context of the valence band maximum, the primary contribution to the total DOS comes from the $p$ orbital of chlorine. However, for the conduction band minimum, the primary contribution to the total DOS is attributed to the $d$ orbital of manganese. Hence, electronic transitions from chlorine ($p$ orbitals) to manganese ($d$ orbitals) are feasible.
\par  A band structure plot visually represents the energy difference between the valence and conduction bands in a material, crucial for understanding its electrical properties and potential applications, such as semiconductor device design and solar cell efficiency optimization. To confirm the indirect bandgap behavior, we plotted the band structure of CsMnCl$_4$, as shown in Fig. \ref{dosbandcollage}\textbf{(b)}. This distinctly illustrates the indirect bandgap behavior of CsMnCl$_4$.
\subsection{Optical Properties}
One can extract the linear optical properties by analyzing the frequency-dependent complex dielectric function, denoted as $\epsilon(\omega)$:
\begin{align}
\epsilon(\omega) =\epsilon_{1}(\omega)+i \epsilon_{2}(\omega).
\end{align}
The dielectric function is expressed through its real and imaginary components, denoted as $\epsilon_{1}(\omega)$ and $\epsilon_{2}(\omega)$, respectively, where $\omega$ signifies the photon frequency. The real component $\epsilon_{1}(\omega)$ is determined using the Kramers–Kr\"{o}nig relationship \cite{toll1956causality}, while the imaginary component $\epsilon_{2}(\omega)$ is computed through momentum matrix elements between valence and conduction wave functions \cite{ehrenreich1959self}. Utilizing $\epsilon_{1}(\omega)$ and $\epsilon_{2}(\omega)$, the refractive index $n(\omega)$ and absorption coefficient $\alpha(\omega)$ can be calculated using the following formulas:
\begin{align}
n(\omega) = \left[\frac{\sqrt{\epsilon^2_{1}+\epsilon^2_{2}}+\epsilon_{1}}{2}\right]^{\frac{1}{2}},\; \mathrm{and}
\end{align}
\begin{align}
\alpha(\omega) =\sqrt{2}\omega\left[\frac{\sqrt{\epsilon^2_{1}+\epsilon^2_{2}}-\epsilon_{1}}{2}\right]^{\frac{1}{2}}.
\end{align}
\par
\begin{figure}[hbt!]
\centering
\includegraphics[width=1.0\linewidth]{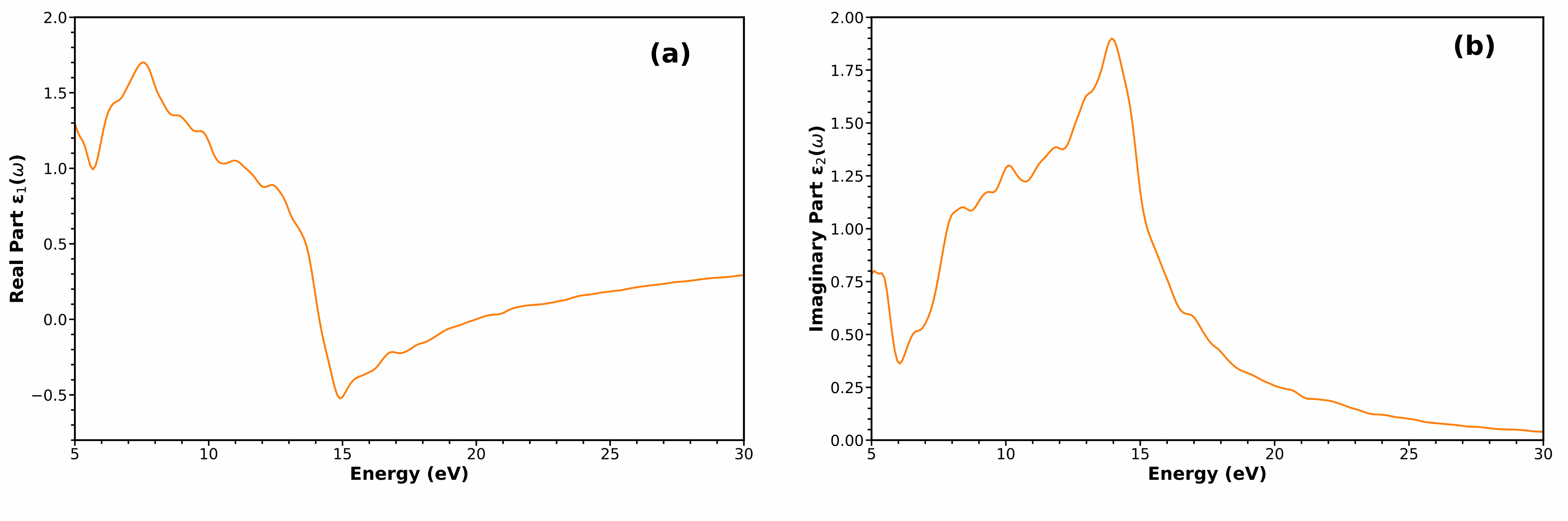}
\caption{\small The PBE calculations yield \textbf{(a)} the real part $\epsilon_{1} (\omega)$ and \textbf{(b)} the imaginary part $\epsilon_{2} (\omega)$ of the complex dielectric function for CsMnCl$_4$.}
\label{Real_Imag}
\end{figure}

 \noindent  
In Fig. (\ref{Real_Imag}), we illustrate the computed $\epsilon_{1}(\omega)$ and $\epsilon_{2}(\omega)$ as functions of $\omega$ for the anticipated compounds as per our predictions. 
\par The real part exhibits its maximum peak around 10 eV, corresponding to $\epsilon_{1}(\omega)=1.7$, in the case of CsMnCl$_4$ in Fig.\ref{Real_Imag}\textbf{(a)}. The static dielectric constants can be determined by evaluating the zero-frequency limits as $\omega$ approaches 0 for $\epsilon_{1}(\omega)$. The static dielectric constant for CsMnCl$_4$ is found to be 1.3. In Fig.\ref{Real_Imag}\textbf{(b)}, the plot of $\epsilon_{2}(\omega)$ indicates that the threshold energy of the dielectric function is approximately 5 eV. This corresponds to the fundamental absorption edge, representing the optical transition between the valence band maximum (VBM) and the conduction band minimum (CBM). In the increasing energy range, the absorptive part of $\epsilon_{2}(\omega)$ reveals a prominent peak at approximately 15 eV. This peak is induced by the transition of Cl$-3p$ electrons to the $s$ states of cations \cite{wang2014structural}.
\begin{figure}[hbt!]
\centering
\includegraphics[width=1.0\linewidth]{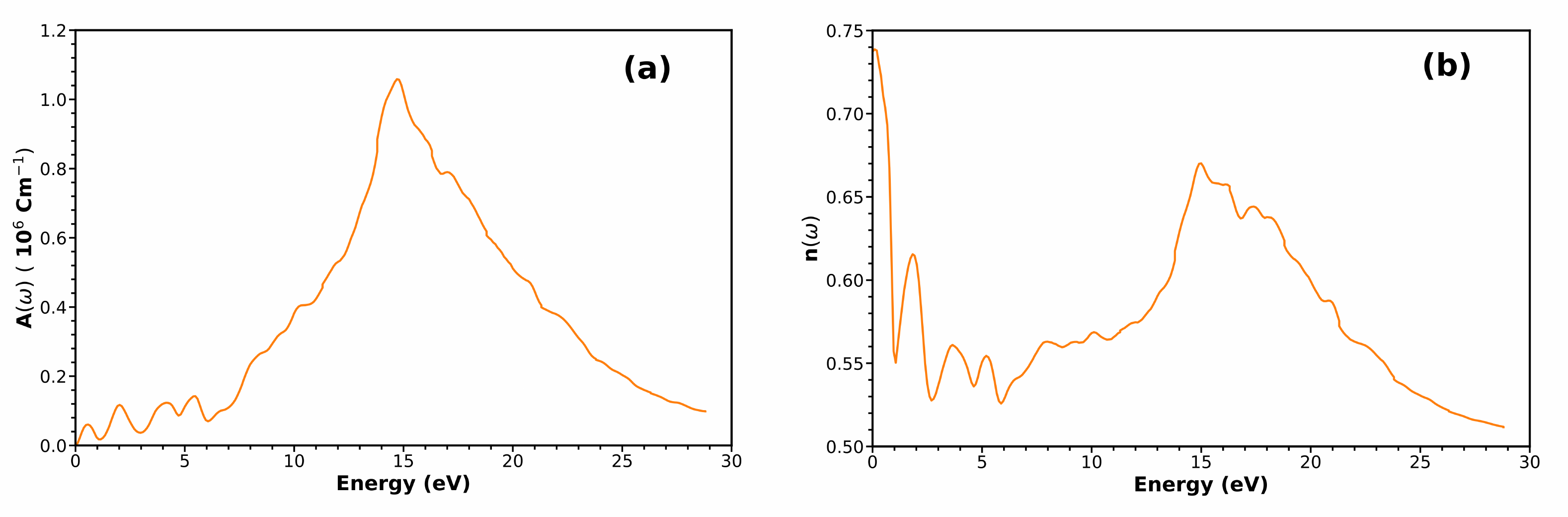}
\caption{\small The PBE calculated \textbf{(a)} absorption $A(\omega)$ and \textbf{(b)} refractive $n(\omega)$ spectrum of CsMnCl$_4$.}
\label{Abs_Refract}
\end{figure}
\par The absorption coefficient $\alpha(\omega)$ characterizes the attenuation of light intensity as it propagates through a material over a unit distance.
As shown in Fig. \ref{Abs_Refract}\textbf{(a)}, the onset of the absorption edge becomes noticeable at approximately 6 eV. This phenomenon is attributed to excited electrons transitioning from Cl$-3p$ states at the upper edge of the valence band to unoccupied cation $3s$ states. It's important to highlight that the absorption coefficient $\alpha(\omega)$ is observed at a value below $6$ eV, specifically in the ultraviolet range. Conversely, the compound CsMnCl$_4$ exhibit significant absorption, attributed to the sharp rise in the absorption coefficient beyond the absorption edge as photon energy surpasses a certain threshold. This characteristic is commonly observed in semiconductors and insulators. Fig.\ref{Abs_Refract}\textbf{(b)} illustrates the measured curve of $n(\omega)$ with respect to photon energy. It is noteworthy that the static refractive index $n(0)$ for incident light is 0.74 in the case of CsMnCl$_4$. Around a photon energy of 0.5 eV, the refractive index $n(\omega)$ reaches its peak value. Subsequently, the energy level gradually diminishes until reaching its minimum point, beyond which there is minimal change in the high-energy zone ($\geq$25 eV).
\section{Conclusion}
The seven materials newly predicted are: CsMnCl$_4$, Rb$_3$Mn$_2$Cl$_9$, Rb$_4$MnCl$_6$, Rb$_3$MnCl$_5$, RbMn$_2$Cl$_7$, RbMn$_4$Cl$_9$, and CsIn$_2$Cl$_7$, predicted through ML and further examined via DFT calculation. We examined the formation energy and constructed the corresponding convex hull plot, leading to the conclusion that the newly identified materials exhibit thermodynamic stability at 0 K. 
All of these discovered compounds demonstrated notable bandgaps and Debye temperatures, suggesting their potential usefulness in optoelectronic devices and applications involving photoluminescence. Among these, CsMnCl$_4$ stands out, with a DFT-PBE bandgap of 1.37 eV, aligning with the Shockley-Queisser limit, rendering it suitable for photovoltaic applications. Furthermore, the optical absorption and emission spectra of CsMnCl$_4$ have been calculated, further supporting their suitability for photovoltaic applications. Moreover, their high Debye temperatures further emphasize their thermal stability, positioning them as promising contenders for practical applications. It is worth mentioning that the magnetic ordering of Mn atoms may moderately change the bandgap and Debye temperature values. However, a detailed study focusing on the magnetic properties of CsMnCl$_4$ is beyond the scope of this work.\\
\par However, our study is not without limitations. The synthetic accessibility of these materials, particularly those with the  Mn$^{3+}$ oxidation state can be challenging due to the complex redox chemistry of Mn, as it exhibits multiple oxidation states ranging from  Mn$^{2+}$ to  Mn$^{7+}$. Achieving the  Mn$^{3+}$ oxidation state (e.g., MnPO$_4$ \cite{huang2014understanding}, MnOCl \cite{feng2023first}, MnOBr \cite{feng2023first}, Mn$_2$O$_3$ \cite{chen2013synthesis}, Mn$_2$S$_3$, Mn$_2$Se$_3$, and Mn$_2$Te$_3$ \cite{liu2024spin} ) often requires precise control over reaction conditions, including temperature, pressure, and stoichiometry.  Additionally, the potential for disproportionation reactions for  Mn$^{3+}$ ions and the unintended introduction of impurities during the synthesis process can lead to the formation of undesirable byproducts or instability in the synthesized compound. In summary,  the synthesis of compounds with the Mn$^{3+}$ oxidation state demands careful experimental design and optimization to overcome these challenges and achieve successful formation while maintaining stability and purity. 
\par Looking forward, the exploration of additional compositional spaces and the incorporation of more complex machine learning models could uncover even more promising candidates for photovoltaic applications. Experimental validation of our computational predictions remains a crucial next step, which will not only test the accuracy of our computational methods but also potentially lead to the development of high-efficiency, lead-free perovskite solar cells. The recent successful discovery of Cs$_3$LuCl$_6$, a novel halide perovskite compound \cite{kim2023high}, which has been demonstrated in white-light-emitting diodes, underscores the potential of data- and ML-driven approaches in material discovery for practical applications. This achievement highlights the efficacy of combining computational predictions with experimental endeavors. Further research will likely deepen our understanding of the mechanisms underpinning the identified trends, thereby enriching our knowledge of halide perovskite materials and guiding future strategies in material design. 
\section*{Acknowledgement}
\noindent This study was supported by National Research Foundation of Korea (NRF-2020M3H4A3081796).
\section*{Author contributions}
\noindent The concept was conceived by Upendra Kumar and Hyeon Woo Kim. Hyeon Woo Kim, Gyanendra Kumar Maurya, Bincy Babu Raj and Sobhit Singh offered valuable suggestions, and Upendra Kumar authored the manuscript, with all authors reading and reviewing it. Ajay Kumar Kushwaha, Sung Beom Cho and Hyunseok Ko supervised the project. 
{
\renewcommand \thesection{S\arabic{section}}
\renewcommand\thetable{S\arabic{table}}
\renewcommand \thefigure{S\arabic{figure}}
\setcounter{section}{0} 
\setcounter{table}{0}   
\setcounter{figure}{0}  

\section*{Supplementary Information}

\section{Structure Predictor code}
\begin{python}
from pymatgen.analysis.structure_prediction.substitutor import Substitutor 
from pymatgen.analysis.structure_prediction.substitution_probability import SubstitutionPredictor
from pymatgen.analysis.structure_matcher import StructureMatcher, ElementComparator
from pymatgen.transformations.standard_transformations import AutoOxiStateDecorationTransformation
from pymatgen.core.periodic_table import Specie, Element
from pymatgen.ext.matproj import MPRester
from pymatgen.core.structure import *
import os
from pprint import pprint
from tqdm import tqdm
###############
mpr = MPRester("------------")
threshold = 0.001 #threshold for substitution/structure predictions
num_subs = 50 # number of highest probability substitutions you wish to see
cation1 = 'Cs' or 'Rb'
cation2 = 'In' or 'Mn' or 'Sn'
possible_charges1 = [1]  # Please check the periodic table!
possible_charges2 = [3]
\end{python}
\begin{python}
for cation1_charge in possible_charges1:  # should be changed [2, 4]
for cation2_charge in possible_charges2:  # should be changed range(2,6)

print(
'oxidation states ({})-({}) is defined'.format(cation1_charge, cation2_charge))

original_species = [Specie(cation1, cation1_charge), Specie(
cation2, cation2_charge), Specie('Cl or Br or I', -1)]

print('SubstitutionPredictor is working ... with {} candidates'.format(num_subs))

subs = SubstitutionPredictor(
threshold=threshold).list_prediction(original_species)
subs.sort(key=lambda x: x['probability'], reverse=True)
subs = subs[0:num_subs]
#pprint(subs)

trial_subs = [list(sub['substitutions'].keys()) for sub in subs]
#pprint(trial_subs)

elem_sys_list = [[specie.element for specie in sub]
for sub in trial_subs]
chemsys_set = set()
for sys in elem_sys_list:
chemsys_set.add("-".join(map(str, sys)))

#pprint(chemsys_set)

# Finding all structures for new chemical systems via Materials API
all_structs = {}
print('Downloading all the structures for ({})-({}) system'.format(cation1_charge, cation2_charge))
for chemsys in tqdm(chemsys_set):
# Getting all structures -- this can take while
all_structs[chemsys] = mpr.get_structures(chemsys)

auto_oxi = AutoOxiStateDecorationTransformation()
oxi_structs = {}

print('Generating oxidation states')
for chemsys in tqdm(all_structs):
oxi_structs[chemsys] = []

for num, struct in enumerate(all_structs[chemsys]):
try:
oxi_structs[chemsys].append({'structure': auto_oxi.apply_transformation(struct),
'id': str(chemsys + "_" + str(num))}
)
except:
continue  # if auto oxidation fails, try next structure

#pprint(oxi_structs)

# Substitute original species into new structures
# Now create a new dictionary trans_structures populated with predicted tructures made up of original species.
# Note: these new predicted structures are TransformedStructure objects

sbr = Substitutor(threshold=threshold)
trans_structs = {}

print('structure transformation...')
for chemsys in tqdm(oxi_structs):
trans_structs[chemsys] = sbr.pred_from_structures(
original_species, oxi_structs[chemsys])

print(trans_structs)

sm = StructureMatcher(
comparator=ElementComparator(), primitive_cell=False)

filtered_structs = {}  # new filtered dictionary
seen_structs = []  # list of all seen structures, independent of chemical system

print("Number of entries BEFORE filtering: " +
str(sum([len(sys) for sys in trans_structs.values()])))

# Remove the duplicated structures
for chemsys in trans_structs:
filtered_structs[chemsys] = []
for struct in trans_structs[chemsys]:
found = False
for struct2 in seen_structs:
if sm.fit(struct.final_structure, struct2.final_structure):
found = True
break
if not found:
filtered_structs[chemsys].append(struct)
seen_structs.append(struct)

print("Number of entries AFTER filtering: " +
str(sum([len(sys) for sys in filtered_structs.values()])))

# Now we wish to run one more filter to remove all duplicate structures already accessible for original system
known_structs = mpr.get_structures(cation1+"-"+cation2+"-"+"Cl")
final_filtered_structs = {}
print("Number of entries BEFORE filtering against MP: " +
str(sum([len(sys) for sys in filtered_structs.values()])))

for chemsys in filtered_structs:
final_filtered_structs[chemsys] = []
for struct in filtered_structs[chemsys]:
found = False
for struct2 in known_structs:
if sm.fit(struct.final_structure, struct2):
found = True
break
if not found:
final_filtered_structs[chemsys].append(struct)

print("Number of entries AFTER filtering against MP: " +
str(sum([len(sys) for sys in final_filtered_structs.values()])))
pprint(final_filtered_structs)

final_structs = {}
for chemsys in final_filtered_structs:
final_structs[chemsys] = [struct.to_snl(
[{"name": "Hyeon Kim", "email": "khw@kicet.re.kr"}]) for struct in final_filtered_structs[chemsys]]

# Start GGA-PBE calculations on the predicted structure file.
b = {}
for key, value in final_structs.items():
if value != []:
b[key] = value

def list_chunk(lst, n):
return [lst[i:i+n] for i in range(0, len(lst), n)]
original_dir=os.getcwd()

for formula in b.keys():
bb = b[formula]
bb_slice = list_chunk(bb, 1)
for num in range(len(bb_slice)):
dict_structure = b[formula][num].as_dict()
dict_from_structure = Structure.from_dict(dict_structure)
dir_name = str(cation1) + str(cation1_charge) + str(cation2) + str(cation2_charge) + "_" + str(formula) + str(num)

if not os.path.exists(dir_name):  # Check if the directory already exists
os.mkdir(dir_name)
os.chdir(dir_name)

dict_from_structure.to(filename=str((formula) + str(num) + ".POSCAR"))
struct = Structure.from_file(filename=str((formula) + str(num) + ".POSCAR"))
os.rename(str(formula) + str(num) + ".POSCAR", "POSCAR")
#                print(os.system("bulk.py"))
#                print(os.system("pbs2.pl"))
#                print(os.system("qsub pbs"))
os.chdir(original_dir)
\end{python}
\section{New Compound Structural Details}
\begin{table}[hbt!]
\centering
\scalebox{1.1}{\begin{tabular}{|c|c|c|c|c|c|c|c|c|c|}
\hline
&&&&&&\\
\quad{} \quad &  \thead{\textbf{\quad Compound \quad}\\ {}} &\thead{\textbf{Lattice }\\ {\textbf{\quad Parameters (\AA) \quad}}}&\thead{\textbf{\quad Lattice \quad \quad}\\ {\textbf{Angle }}} & \thead{\textbf{\quad Space Group \quad}\\ {\textbf{Type}}}  & \thead{\textbf{Crystal }\\ {\textbf{Structure}}}&\thead{\quad \textbf{Space Group} \quad\\ {\quad \textbf{Number}}}\quad \\
&&&&&&\\
\hline
&&&&&&\\
\quad \thead{ {} \\ 1  \quad \quad \\{}}  &     \quad \thead{ {} \\ CsMnCl$_4$ \\ {}} &  \quad \thead{a = 7.37 \AA \\
b = 9.92 \AA\\
c = 9.97 \AA} \quad      &  \quad \thead{$\alpha$ = 90$^{\circ}$ \\
$\beta$ =  91.98$^{\circ}$\\
$\gamma$ = 90$^{\circ}$ } \quad    &   \quad \thead{ {} \\  \textsl{P2$_1$/c} \\ {}} &    \quad \thead{ {} \\   Monoclinic \\ {} \quad }& \quad \thead{ {} \\ 14 \\ {}}\\
&&&&&&\\
\hline
&&&&&&\\
\quad \thead{ {} \\  2  \quad \quad \\ {}}  &   \quad \thead{ {} \\ Rb$_3$Mn$_2$Cl$_9$ \\ {}} & \quad \thead{a = 7.13 \AA   \\
b = 7.13 \AA \\
c = 18.23 \AA } \quad  &   \quad \thead{$\alpha$ = 90$^{\circ}$ \\
$\beta$ = 90$^{\circ}$  \\
$\gamma$ = 120$^{\circ}$ } \quad    &   \quad \thead{ {} \\  \textsl{P6$_3$/mmc} \\ {}} &    \quad \thead{ {} \\   Hexagonal \\ {} }&\quad \thead{ {} \\  194 \\ {} }\\
&&&&&&\\
\hline 
&&&&&&\\
\quad \thead{ {} \\ 3  \quad \quad \\ {}}  &   \quad \thead{ {} \\  Rb$_4$MnCl$_6$ \\ {}} &  \quad \thead{a = 8.96 \AA  \\
b = 8.96 \AA  \\
c = 8.96 \AA } \quad     & \quad \thead{$\alpha$ = 89.22$^{\circ}$ \\
$\beta$ = 89.22$^{\circ}$ \\
$\gamma$ = 89.22$^{\circ}$ } \quad       &    \quad \thead{ {} \\  \textsl{R$\bar{3}$c} \\ {}} &     \quad \thead{ {} \\  Trigonal \\ {}} &\quad \thead{ {} \\  167 \\ {}} \\
&&&&&&\\
\hline 
&&&&&&\\
\quad \thead{ {} \\ 4 \quad \quad \\ {}}  &   \quad \thead{ {} \\  Rb$_3$MnCl$_5$ \\ {}}& \quad \thead{a = 9.03 \AA  \\
b = 9.03 \AA  \\
c = 14.62 \AA } \quad    &  \quad \thead{$\alpha$ = 90$^{\circ}$  \\
$\beta$ = 90$^{\circ}$ \\
$\gamma$ = 90$^{\circ}$ } \quad    &   \quad \thead{ {} \\  \textsl{I4/mcm} \\ {}} &   \quad \thead{ {} \\    Tetragonal \\ {}} & \quad \thead{ {} \\ 140 \\ {}} \\
&&&&&&\\
\hline 
&&&&&&\\
\quad \thead{ {} \\  5  \quad \quad \\ {}}  & \quad \thead{ {} \\    RbMn$_2$Cl$_7$ \\ {}} & \quad \thead{a = 7.06 \AA  \\
b = 11.23 \AA \\
c = 7.45 \AA } \quad     &  \quad \thead{$\alpha$ = 90$^{\circ}$ \\
$\beta$ = 104$^{\circ}$  \\
$\gamma$ = 90$^{\circ}$  } \quad    &   \quad \thead{ {} \\  \textsl{P2$_1$/m} \\ {}} &    \quad \thead{ {} \\   Monoclinic \\ {}} & \quad \thead{ {} \\ 11 \\ {}}\\
&&&&&&\\
\hline 
&&&&&&\\
\quad \thead{ {} \\ 6 \quad \quad \\ {}}  &\quad \thead{ {} \\     RbMn$_4$Cl$_9$ \\ {}} &  \quad \thead{a = 11.77 \AA  \\
b = 11.77 \AA \\
c = 10.42 \AA } \quad    &  \quad \thead{$\alpha$ = 90$^{\circ}$  \\
$\beta$ = 90$^{\circ}$ \\
$\gamma$ = 90$^{\circ}$ } \quad    &   \quad \thead{ {} \\  \textsl{I4$_1$/a} \\ {}} &   \quad \thead{ {} \\    Tetragonal \\ {}} & \quad \thead{ {} \\ 88 \\ {}}\\
&&&&&&\\
\hline
&&&&&&\\
\quad \thead{ {} \\ 7  \quad \quad \\ {}}  &  \quad \thead{ {} \\    CsIn$_2$Cl$_7$ \\ {}}& \quad \thead{a = 7.58 \AA  \\
b = 6.51 \AA \\
c = 12.75 \AA } \quad       &   \quad \thead{$\alpha$ = 90$^{\circ}$ \\
$\beta$ = 90.89$^{\circ}$ \\
$\gamma$ = 90$^{\circ}$  } \quad   &   \quad \thead{ {} \\ \textsl{P2/c} \\ {}} &   \quad \thead{ {} \\    Monoclinic \\ {}} & \quad \thead{ {} \\ 13 \\ {}} \\
&&&&&&\\
\hline
\end{tabular}}
\caption{\small The structural intricacies of newly forecasted compounds.}
\label{New_Compound}
\end{table}
\begin{figure}[hbt!]
\centering
\includegraphics[width=.90\linewidth]{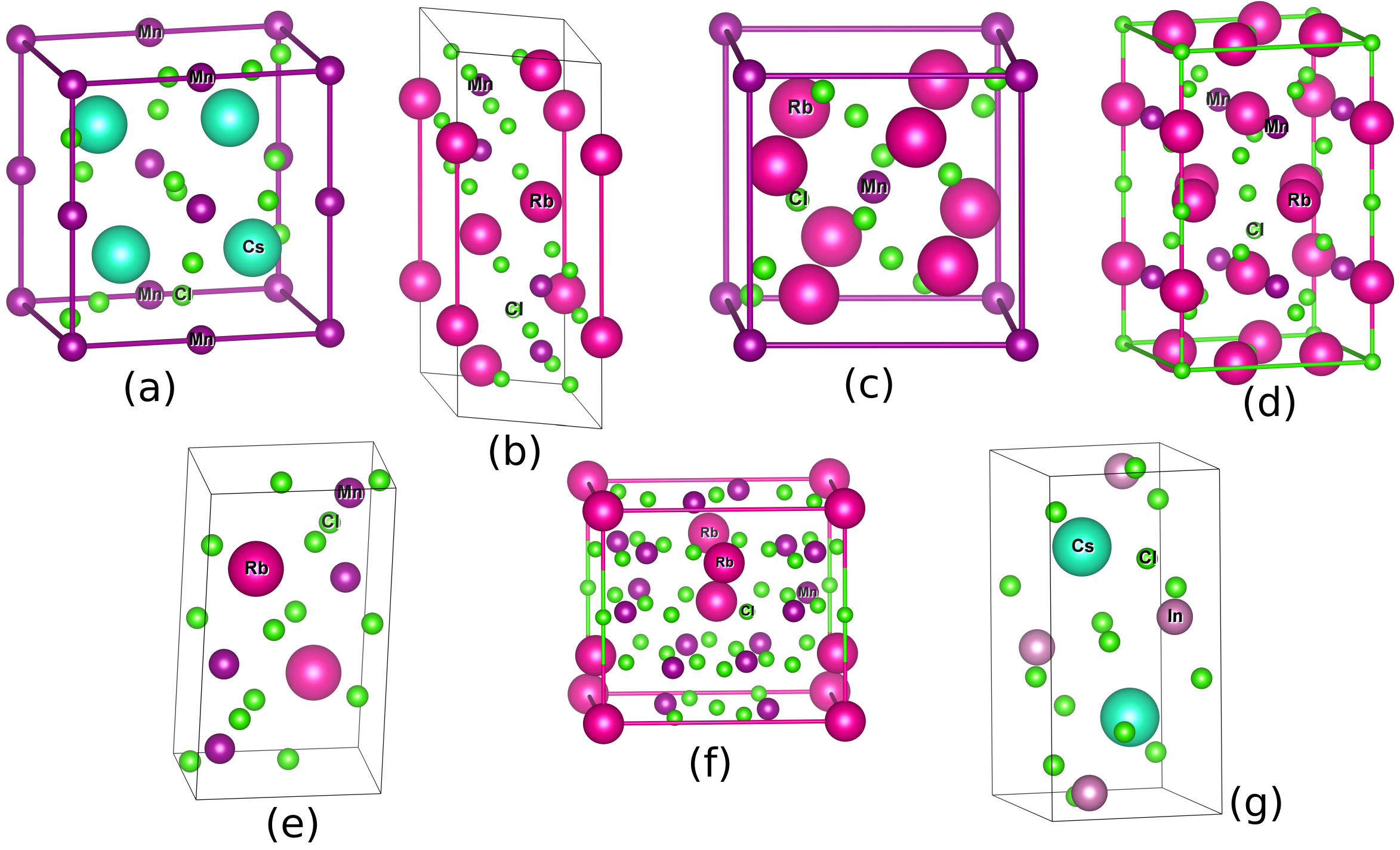}
\caption{\small Crystal structures of \textbf{(a)}  CsMnCl$_4$ (\textsl{P2$_1$/c}), \textbf{(b)} Rb$_3$Mn$_2$Cl$_9$ (\textsl{P6$_3$/mmc}), \textbf{(c)} Rb$_4$MnCl$_6$ (\textsl{R$\bar{3}$c}), \textbf{(d)} Rb$_3$MnCl$_5$ (\textsl{I4/mcm}), \textbf{(e)} RbMn$_2$Cl$_7$ (\textsl{P2$_1$/m}), \textbf{(f)} RbMn$_4$Cl$_9$ (\textsl{I4$_1$/a}) ,and \textbf{(g)} CsIn$_2$Cl$_7$ (\textsl{P2/c}).}
\label{ball_stick}
\end{figure}

\section{Convex Hull}
\begin{table}[hbt!]
\centering
\scalebox{0.85}{
\begin{tabular}{|c|c|c|c|c|c|c|}
\hline
&&&&&& \\
\quad {} \quad & \textbf{Composition} &        \textbf{Space Group} & \textbf{Crystal System} &       \textbf{Ehull (eV/atom)} & \thead{\textbf{ Band Gap(eV)} } &  \textbf{Direct} \\
& &&&&& \\
\hline
& &&&&& \\
0 &     Cs$_3$InI$_6$ &     (\textsl{Fm$\bar{3}$m}, 225) &          cubic &   0.00 &      1.59 &    True \\
1 &      CsInI$_4$ &       (\textsl{R3c}, 161) &       trigonal &   0.00 &      2.77 &    True \\
2 &    Cs$_3$In$_2$I$_9$ &  (\textsl{P6$_3$/mmc}, 194) &      hexagonal &   0.01 &      1.17 &    True \\
3 &      CsInI$_4$ &     (\textsl{P2$_1$/c}, 14) &     monoclinic &   0.01 &      2.15 &   False \\
4 &    Cs$_3$In$_2$I$_9$ &      (\textsl{R$\bar{3}$c}, 167) &       trigonal &   0.01 &      1.48 &    True \\
5 &      CsInI$_4$ &       (\textsl{Pnma}, 62) &   orthorhombic &   0.02 &      2.18 &    True \\
& &&&&& \\
\hline
\end{tabular}
}
\caption{\small \small Predicted structures in the  Cs-In-I ternary compound family.}
\end{table}
\begin{table}[hbt!]
\centering
\scalebox{0.85}{
\begin{tabular}{|c|c|c|c|c|c|c|}
\hline
&&&&&& \\
\quad {} \quad & \textbf{Composition} &        \textbf{Space Group} & \textbf{Crystal System} &       \textbf{Ehull (eV/atom)} & \thead{\textbf{ Band Gap(eV)} } &  \textbf{Direct} \\
& &&&&& \\
\hline
& &&&&& \\
0  &    Cs$_3$MnCl$_6$ &       (\textsl{C2/c}, 15) &     monoclinic &   0.00 &      0.16 &   False \\
1  &    Cs$_2$MnCl$_5$ &    (\textsl{P4/mmm}, 123) &     tetragonal &   0.00 &      0.16 &    True \\

2  &    CsMn$_2$Cl$_7$ &     (\textsl{P2$_1$/m}, 11) &     monoclinic &   0.00 &      0.76 &   False \\
3  &    Cs$_4$MnCl$_6$ &      (\textsl{R$\bar{3}$c}, 167) &       trigonal &   0.00 &      2.66 &    True \\
4  &     CsMnCl$_3$ &  (\textsl{P6$_3$/mmc}, 194) &      hexagonal &   0.00 &      1.39 &    True \\
5  &    Cs$_2$MnCl$_4$ &     (\textsl{Pna2$_1$}, 33) &   orthorhombic &   0.00 &      2.85 &    True \\
6  &    Cs$_2$MnCl$_4$ &       (\textsl{Pnma}, 62) &   orthorhombic &   0.00 &      2.79 &    True \\
7  &     CsMnCl$_3$ &    (\textsl{P6$_1$22}, 178) &      hexagonal &   0.00 &      1.39 &    True \\
8  &   Cs$_3$Mn$_2$Cl$_9$ &  (\textsl{P6$_3$/mmc}, 194) &      hexagonal &   0.00 &      0.00 &   False \\
9  &    CsMnCl$_4$ &     (\textsl{P2$_1$/c}, 14) &     monoclinic &   0.00 &      1.37 &   False\\
10 &   Cs$_3$Mn$_2$Cl$_9$ &     (\textsl{P$\bar{3}$m1}, 164) &       trigonal &   0.01 &      0.10 &   False \\
11 &   Cs$_3$Mn$_2$Cl$_9$ &      (\textsl{R$\bar{3}$c}, 167) &       trigonal &   0.01 &      0.00 &   False \\
12 &  CsMn$_2$Cl$_7$ &       (\textsl{P2/c}, 13) &     monoclinic &   0.01 &      0.42 &   False     \\
13 &     CsMnCl$_3$ &  (\textsl{P6$_3$/mmc}, 194) &      hexagonal &   0.01 &      1.48 &    True \\
14 &     CsMnCl$_4$ &       (\textsl{Pnma}, 62) &   orthorhombic &   0.02 &      0.61 &    True \\
15 &     CsMnCl$_3$ &  (\textsl{P6$_3$/mmc}, 194) &      hexagonal &   0.02 &      1.46 &   False \\
16 &  Cs$_4$Mn$_3$Cl$_{10}$ &       (\textsl{Cmce}, 64) &   orthorhombic &   0.02 &      1.75 &    True \\
17 &     CsMnCl$_4$ &       (\textsl{P4/n}, 85) &     tetragonal &   0.02 &      1.33 &   False \\
18 &  Cs$_7$Mn$_4$Cl$_{15}$ &     (\textsl{P2$_1$/c}, 14) &     monoclinic &   0.03 &      1.91 &    True \\
19 &   Cs$_3$Mn$_2$Cl$_7$ &    (\textsl{I4/mmm}, 139) &     tetragonal &   0.03 &      1.43 &   False \\
20 &    Cs$_3$MnCl$_6$ &     (\textsl{Fm$\bar{3}$m}, 225) &          cubic &   0.04 &      0.00 &   False \\
21 &    Cs$_4$MnCl$_6$ &     (\textsl{Im$\bar{3}$m}, 229) &          cubic &   0.04 &      2.61 &    True \\
22 &     CsMnCl$_3$ &    (\textsl{I4/mcm}, 140) &     tetragonal &   0.04 &      1.32 &    True \\
23 &     CsMnCl$_3$ &       (\textsl{Pnma}, 62) &   orthorhombic &   0.04 &      1.54 &    True \\
24 &     CsMnCl$_3$ &       (\textsl{Pnma}, 62) &   orthorhombic &   0.04 &      2.01 &   False \\
25 &     CsMnCl$_3$ &     (\textsl{Pm$\bar{3}$m}, 221) &          cubic &   0.05 &      1.56 &   False \\
26 &    Cs$_4$MnCl$_6$ &       (\textsl{C2/c}, 15) &     monoclinic &   0.12 &      1.30 &    True \\
27 &   Cs$_3$Mn$_2$Cl$_7$ &    (\textsl{I4/mmm}, 139) &     tetragonal &   0.13 &      1.20 &    True \\
& &&&&& \\
\hline
\end{tabular}
}
\caption{\small \small Predicted structures in the  Cs-Mn-Cl ternary compound family.}
\end{table}
\begin{table}[hbt!]
\centering
\scalebox{0.85}{
\begin{tabular}{|c|c|c|c|c|c|c|}
\hline
&&&&&& \\
\quad {} \quad& \textbf{Composition} &        \textbf{Space Group} & \textbf{Crystal System} &       \textbf{Ehull (eV/atom)} & \thead{\textbf{ Band Gap(eV)} } &  \textbf{Direct} \\
& &&&&& \\
\hline
& &&&&& \\
0 &      CsSnI$_3$ &     (\textsl{Pnma}, 62) &   orthorhombic &   0.00 &      2.07 &   False \\
1 &     Cs$_4$SnI$_6$ &    (R$\bar{3}$c, 167) &       trigonal &   0.00 &      3.14 &    True \\
2 &     CsSn$_2$I$_5$ &  (\textsl{I4/mcm}, 140) &     tetragonal &   0.00 &      1.65 &   False \\
3 &     Cs$_2$SnI$_6$ &   (\textsl{Fm$\bar{3}$m}, 225) &          cubic &   0.00 &      0.46 &    True \\
4 &     Cs$_2$SnI$_6$ &   (\textsl{P2$_1$/c}, 14) &     monoclinic &   0.00 &      0.62 &    True \\
5 &      CsSnI$_3$ &     (\textsl{Pnma}, 62) &   orthorhombic &   0.00 &      0.86 &    True \\
6 &      CsSnI$_3$ &   (\textsl{Pm$\bar{3}$m}, 221) &          cubic &   0.01 &      1.78 &    True \\
7 &      CsSnI$_3$ &   (\textsl{P2$_1$/c}, 14) &     monoclinic &   0.01 &      2.16 &    True \\
& &&&&& \\
\hline
\end{tabular}
}
\caption{\small \small Predicted structures in the  Cs-Sn-I ternary compound family.}
\end{table}
\begin{table}[hbt!]
\centering
\scalebox{0.75}{
\begin{tabular}{|c|c|c|c|c|c|c|}
\hline
&&&&&& \\
\quad {} \quad& \textbf{Composition} &        \textbf{Space Group} & \textbf{Crystal System} &       \textbf{Ehull (eV/atom)} & \thead{\textbf{ Band Gap(eV)} } &  \textbf{Direct} \\
& &&&&& \\
\hline
& &&&&& \\
0  &   Rb$_3$Mn$_2$Cl$_9$ &  (\textsl{P6$_3$/mmc}, 194) &      hexagonal &   0.00 &      0.00 &   False \\
1  &    Rb$_2$MnCl$_4$ &          (Cc, 9) &     monoclinic &   0.00 &      2.79 &    True \\
2  &    Rb$_4$MnCl$_6$ &      (\textsl{R$\bar{3}$c}, 167) &       trigonal &   0.00 &      2.61 &    True \\
3  &    Rb$_3$MnCl$_5$ &    (\textsl{I4/mcm}, 140) &     tetragonal &   0.00 &      2.84 &   False \\
4  &    Rb$_2$MnCl$_5$ &    (\textsl{P4/mmm}, 123) &     tetragonal &   0.00 &      0.00 &   False \\
5  &    RbMn$_2$Cl$_7$ &     (\textsl{P2$_1$/m}, 11) &     monoclinic &   0.00 &      0.79 &   False \\
6  &     RbMnCl$_3$ &    (\textsl{P6$_3$cm}, 185) &      hexagonal &   0.00 &      1.37 &    True \\
7  &    Rb$_3$MnCl$_6$ &       (\textsl{C2/c}, 15) &     monoclinic &   0.00 &      0.14 &   False \\
8  &    RbMn$_4$Cl$_9$ &     (\textsl{I4$_1$/a}, 88) &     tetragonal &   0.00 &      1.89 &   False \\
9  &   Rb$_3$Mn$_2$Cl$_9$ &     (\textsl{P$\bar{3}$m1}, 164) &       trigonal &   0.00 &      0.07 &   False \\
10 &    Rb$_2$MnCl$_4$ &     (\textsl{Pna2$_1$}, 33) &   orthorhombic &   0.00 &      2.84 &    True \\
11 &    Rb$_2$MnCl$_4$ &     (\textsl{P2$_1$/c}, 14) &     monoclinic &   0.00 &      2.88 &   False \\
12 &     RbMnCl$_3$ &       (\textsl{C2/c}, 15) &     monoclinic &   0.00 &      1.45 &   False \\
13 &    Rb$_2$MnCl$_4$ &     (\textsl{P2$_1$/m}, 11) &     monoclinic &   0.00 &      2.72 &   False \\
14 &     RbMnCl$_3$ &       (\textsl{Cmcm}, 63) &   orthorhombic &   0.00 &      1.41 &    True \\
15 &     RbMnCl$_3$ &  (\textsl{P6$_3$/mmc}, 194) &      hexagonal &   0.00 &      1.29 &    True \\
16 &     RbMnCl$_3$ &  (\textsl{P6$_3$/mmc}, 194) &      hexagonal &   0.00 &      1.29 &    True \\
17 &    Rb$_2$MnCl$_4$ &       (\textsl{Pnma}, 62) &   orthorhombic &   0.00 &      2.75 &    True \\
18 &    Rb$_2$MnCl$_4$ &     (\textsl{Pna2$_1$}, 33) &   orthorhombic &   0.00 &      2.76 &    True \\
19 &     RbMnCl$_3$ &      (\textsl{R$\bar{3}$m}, 166) &       trigonal &   0.01 &      1.35 &   False \\
20 &     RbMnCl$_3$ &  (\textsl{P6$_3$/mmc}, 194) &      hexagonal &   0.01 &      1.34 &    True \\
21 &   Rb$_3$Mn$_2$Cl$_9$ &      (\textsl{R$\bar{3}$c}, 167) &       trigonal &   0.01 &      0.00 &   False \\
22 &    RbMn$_2$Cl$_7$ &       (\textsl{P2/c}, 13) &     monoclinic &   0.01 &      0.43 &   False \\
23 &     RbMnCl$_4$ &     (\textsl{P2$_1$/c}, 14) &     monoclinic &   0.01 &      1.34 &   False \\
24 &     RbMnCl$_3$ &       (\textsl{Pnma}, 62) &   orthorhombic &   0.02 &      1.97 &   False \\
25 &     RbMnCl$_4$ &       (\textsl{P4/n}, 85) &     tetragonal &   0.02 &      1.36 &   False \\
26 &   Rb$_3$Mn$_2$Cl$_7$ &    (\textsl{I4/mmm}, 139) &     tetragonal &   0.02 &      1.41 &    True \\
27 &     RbMnCl$_3$ &      (\textsl{R$\bar{3}$c}, 167) &       trigonal &   0.02 &      1.36 &   False \\
28 &     RbMnCl$_4$ &       (\textsl{Pnma}, 62) &   orthorhombic &   0.02 &      0.78 &    True \\
29 &     RbMnCl$_3$ &       (\textsl{Pnma}, 62) &   orthorhombic &   0.02 &      1.44 &    True \\
30 &     RbMnCl$_3$ &       (\textsl{Pnma}, 62) &   orthorhombic &   0.02 &      1.45 &    True \\
31 &    Rb$_2$MnCl$_4$ &    (\textsl{I4/mmm}, 139) &     tetragonal &   0.02 &      1.28 &    True \\
32 &     RbMnCl$_3$ &     (\textsl{Pm$\bar{3}$m}, 221) &          cubic &   0.03 &      1.42 &   False \\
33 &    Rb$_4$MnCl$_6$ &     (\textsl{Im$\bar{3}$m}, 229) &          cubic &   0.04 &      2.52 &    True \\
34 &    Rb$_4$MnCl$_6$ &     (\textsl{Im$\bar{3}$m}, 229) &          cubic &   0.04 &      2.52 &    True \\
35 &    Rb$_3$MnCl$_6$ &     (\textsl{Fm$\bar{3}$m}, 225) &          cubic &   0.05 &      0.00 &   False \\
36 &    Rb$_4$MnCl$_6$ &       (\textsl{C2/c}, 15) &     monoclinic &   0.12 &      1.09 &    True \\
& &&&&& \\
\hline
\end{tabular}
}
\caption{\small \small Predicted structures in the  Rb-Mn-Cl ternary compound family.}
\end{table}
\begin{table}[hbt!]
\centering
\scalebox{0.85}{\begin{tabular}{|c|c|c|c|c|c|c|}
\hline
&&&&&& \\
\quad {} \quad&              \textbf{Composition} &        \textbf{Space Group} & \textbf{Crystal System} &       \textbf{Ehull (eV/atom)} & \thead{\textbf{ Band Gap(eV)} } &  \textbf{Direct} \\
&&&&&& \\
\hline
&&&&&& \\
0  &      CsInCl$_4$ &       (\textsl{Pnma}, 62) &   orthorhombic &   0.00 &      3.83 &    True \\
1  &     Cs$_3$InCl$_6$ &       (\textsl{C2/c}, 15) &     monoclinic &   0.00 &      3.67 &    True \\
2  &     CsIn$_2$Cl$_7$ &       (\textsl{P2/c}, 13) &     monoclinic &   0.00 &      3.01 &   False \\
3  &    Cs$_3$In$_2$Cl$_9$ &  (\textsl{P6$_3$/mmc}, 194) &      hexagonal &   0.00 &      3.52 &    True \\
4  &    Cs$_3$In$_2$Cl$_9$ &    (\textsl{P6$_3$cm}, 185) &      hexagonal &   0.00 &      3.42 &    True \\
5  &     CsIn$_2$Cl$_7$ &     (\textsl{P2$_1$/m}, 11) &     monoclinic &   0.00 &      3.30 &   False \\
6  &     Cs$_3$InCl$_6$ &     (\textsl{P2$_1$/c}, 14) &     monoclinic &   0.00 &      3.55 &    True \\
7  &      CsInCl$_4$ &     (\textsl{P2$_1/c$}, 14) &     monoclinic &   0.00 &      3.79 &   False \\
8  &     Cs$_2$InCl$_5$ &       (\textsl{Pnma}, 62) &   orthorhombic &   0.01 &      2.83 &    True \\
9  &    Cs$_3$In$_2$Cl$_9$ &     (\textsl{P$\bar{3}$m1}, 164) &       trigonal &   0.03 &      1.86 &   False \\
10 &  Cs$_{11}$In$_6$Cl$_{29}$ &       (\textsl{C2/m}, 12) &     monoclinic &   0.04 &      2.80 &   False \\
11 &     CsIn$_2$Cl$_7$ &       (\textsl{Pnma}, 62) &   orthorhombic &   0.09 &      2.74 &    True \\
&&&&&& \\
\hline
\end{tabular}}
\caption{\small Predicted structures in the  Cs-In-Cl ternary compound family.}
\end{table}

}
\clearpage 

\begin{thebibliography}{35}%
\makeatletter
\providecommand \@ifxundefined [1]{%
\@ifx{#1\undefined}
}%
\providecommand \@ifnum [1]{%
\ifnum #1\expandafter \@firstoftwo
\else \expandafter \@secondoftwo
\fi
}%
\providecommand \@ifx [1]{%
\ifx #1\expandafter \@firstoftwo
\else \expandafter \@secondoftwo
\fi
}%
\providecommand \natexlab [1]{#1}%
\providecommand \enquote  [1]{``#1''}%
\providecommand \bibnamefont  [1]{#1}%
\providecommand \bibfnamefont [1]{#1}%
\providecommand \citenamefont [1]{#1}%
\providecommand \href@noop [0]{\@secondoftwo}%
\providecommand \href [0]{\begingroup \@sanitize@url \@href}%
\providecommand \@href[1]{\@@startlink{#1}\@@href}%
\providecommand \@@href[1]{\endgroup#1\@@endlink}%
\providecommand \@sanitize@url [0]{\catcode `\\12\catcode `\$12\catcode
`\&12\catcode `\#12\catcode `\^12\catcode `\_12\catcode `\%12\relax}%
\providecommand \@@startlink[1]{}%
\providecommand \@@endlink[0]{}%
\providecommand \url  [0]{\begingroup\@sanitize@url \@url }%
\providecommand \@url [1]{\endgroup\@href {#1}{\urlprefix }}%
\providecommand \urlprefix  [0]{URL }%
\providecommand \Eprint [0]{\href }%
\providecommand \doibase [0]{https://doi.org/}%
\providecommand \selectlanguage [0]{\@gobble}%
\providecommand \bibinfo  [0]{\@secondoftwo}%
\providecommand \bibfield  [0]{\@secondoftwo}%
\providecommand \translation [1]{[#1]}%
\providecommand \BibitemOpen [0]{}%
\providecommand \bibitemStop [0]{}%
\providecommand \bibitemNoStop [0]{.\EOS\space}%
\providecommand \EOS [0]{\spacefactor3000\relax}%
\providecommand \BibitemShut  [1]{\csname bibitem#1\endcsname}%
\let\auto@bib@innerbib\@empty
\bibitem [{\citenamefont {Yin}\ \emph {et~al.}(2015)\citenamefont {Yin},
\citenamefont {Yang}, \citenamefont {Kang}, \citenamefont {Yan},\ and\
\citenamefont {Wei}}]{yin2015halide}%
\BibitemOpen
\bibfield  {author} {\bibinfo {author} {\bibfnamefont {W.-J.}\ \bibnamefont
{Yin}}, \bibinfo {author} {\bibfnamefont {J.-H.}\ \bibnamefont {Yang}},
\bibinfo {author} {\bibfnamefont {J.}~\bibnamefont {Kang}}, \bibinfo {author}
{\bibfnamefont {Y.}~\bibnamefont {Yan}},\ and\ \bibinfo {author}
{\bibfnamefont {S.-H.}\ \bibnamefont {Wei}},\ }\bibfield  {title} {\bibinfo
{title} {Halide perovskite materials for solar cells: a theoretical review},\
}\href {https://doi.org/10.1039/D3TA06988E} {\bibfield  {journal} {\bibinfo
{journal} {Journal of Materials Chemistry A}\ }\textbf {\bibinfo {volume}
{3}},\ \bibinfo {pages} {8926} (\bibinfo {year} {2015})}\BibitemShut
{NoStop}%
\bibitem [{\citenamefont {Unger}\ \emph {et~al.}(2017)\citenamefont {Unger},
\citenamefont {Kegelmann}, \citenamefont {Suchan}, \citenamefont
{S{\"o}rell}, \citenamefont {Korte},\ and\ \citenamefont
{Albrecht}}]{unger2017roadmap}%
\BibitemOpen
\bibfield  {author} {\bibinfo {author} {\bibfnamefont {E.}~\bibnamefont
{Unger}}, \bibinfo {author} {\bibfnamefont {L.}~\bibnamefont {Kegelmann}},
\bibinfo {author} {\bibfnamefont {K.}~\bibnamefont {Suchan}}, \bibinfo
{author} {\bibfnamefont {D.}~\bibnamefont {S{\"o}rell}}, \bibinfo {author}
{\bibfnamefont {L.}~\bibnamefont {Korte}},\ and\ \bibinfo {author}
{\bibfnamefont {S.}~\bibnamefont {Albrecht}},\ }\bibfield  {title} {\bibinfo
{title} {Roadmap and roadblocks for the band gap tunability of metal halide
perovskites},\ }\href {https://doi.org/10.1039/C7TA00404D} {\bibfield
{journal} {\bibinfo  {journal} {Journal of Materials Chemistry A}\ }\textbf
{\bibinfo {volume} {5}},\ \bibinfo {pages} {11401} (\bibinfo {year}
{2017})}\BibitemShut {NoStop}%
\bibitem [{\citenamefont {Lei}\ \emph {et~al.}(2021)\citenamefont {Lei},
\citenamefont {Dong}, \citenamefont {Gundogdu},\ and\ \citenamefont
{So}}]{lei2021metal}%
\BibitemOpen
\bibfield  {author} {\bibinfo {author} {\bibfnamefont {L.}~\bibnamefont
{Lei}}, \bibinfo {author} {\bibfnamefont {Q.}~\bibnamefont {Dong}}, \bibinfo
{author} {\bibfnamefont {K.}~\bibnamefont {Gundogdu}},\ and\ \bibinfo
{author} {\bibfnamefont {F.}~\bibnamefont {So}},\ }\bibfield  {title}
{\bibinfo {title} {Metal halide perovskites for laser applications},\ }\href
{https://doi.org/10.1002/adfm.202010144} {\bibfield  {journal} {\bibinfo
{journal} {Advanced Functional Materials}\ }\textbf {\bibinfo {volume}
{31}},\ \bibinfo {pages} {2010144} (\bibinfo {year} {2021})}\BibitemShut
{NoStop}%
\bibitem [{\citenamefont {Shrestha}\ \emph {et~al.}(2022)\citenamefont
{Shrestha}, \citenamefont {Li}, \citenamefont {Tsai}, \citenamefont {Hou},
\citenamefont {Huang}, \citenamefont {Ghosh}, \citenamefont {Shyue},
\citenamefont {Wang}, \citenamefont {Tretiak}, \citenamefont {Ma} \emph
{et~al.}}]{shrestha2022long}%
\BibitemOpen
\bibfield  {author} {\bibinfo {author} {\bibfnamefont {S.}~\bibnamefont
{Shrestha}}, \bibinfo {author} {\bibfnamefont {X.}~\bibnamefont {Li}},
\bibinfo {author} {\bibfnamefont {H.}~\bibnamefont {Tsai}}, \bibinfo {author}
{\bibfnamefont {C.-H.}\ \bibnamefont {Hou}}, \bibinfo {author} {\bibfnamefont
{H.-H.}\ \bibnamefont {Huang}}, \bibinfo {author} {\bibfnamefont
{D.}~\bibnamefont {Ghosh}}, \bibinfo {author} {\bibfnamefont {J.-J.}\
\bibnamefont {Shyue}}, \bibinfo {author} {\bibfnamefont {L.}~\bibnamefont
{Wang}}, \bibinfo {author} {\bibfnamefont {S.}~\bibnamefont {Tretiak}},
\bibinfo {author} {\bibfnamefont {X.}~\bibnamefont {Ma}}, \emph {et~al.},\
}\bibfield  {title} {\bibinfo {title} {Long carrier diffusion length in
two-dimensional lead halide perovskite single crystals},\ }\href
{https://doi.org/10.1016/j.chempr.2022.01.008} {\bibfield  {journal}
{\bibinfo  {journal} {Chem}\ }\textbf {\bibinfo {volume} {8}},\ \bibinfo
{pages} {1107} (\bibinfo {year} {2022})}\BibitemShut {NoStop}%
\bibitem [{\citenamefont {Jathar}\ \emph {et~al.}(2021)\citenamefont {Jathar},
\citenamefont {Rondiya}, \citenamefont {Bade}, \citenamefont {Nasane},
\citenamefont {Barma}, \citenamefont {Jadhav}, \citenamefont {Rokade},
\citenamefont {Kore}, \citenamefont {Nilegave}, \citenamefont {Funde} \emph
{et~al.}}]{jathar2021facile}%
\BibitemOpen
\bibfield  {author} {\bibinfo {author} {\bibfnamefont {S.~B.}\ \bibnamefont
{Jathar}}, \bibinfo {author} {\bibfnamefont {S.~R.}\ \bibnamefont {Rondiya}},
\bibinfo {author} {\bibfnamefont {B.~R.}\ \bibnamefont {Bade}}, \bibinfo
{author} {\bibfnamefont {M.~P.}\ \bibnamefont {Nasane}}, \bibinfo {author}
{\bibfnamefont {S.~V.}\ \bibnamefont {Barma}}, \bibinfo {author}
{\bibfnamefont {Y.~A.}\ \bibnamefont {Jadhav}}, \bibinfo {author}
{\bibfnamefont {A.~V.}\ \bibnamefont {Rokade}}, \bibinfo {author}
{\bibfnamefont {K.~B.}\ \bibnamefont {Kore}}, \bibinfo {author}
{\bibfnamefont {D.~S.}\ \bibnamefont {Nilegave}}, \bibinfo {author}
{\bibfnamefont {A.~M.}\ \bibnamefont {Funde}}, \emph {et~al.},\ }\bibfield
{title} {\bibinfo {title} {Facile method for synthesis of cspbbr3 perovskite
at room temperature for solar cell applications},\ }\href
{http://dx.doi.org/10.30919/esmm5f1036} {\bibfield  {journal} {\bibinfo
{journal} {ES Materials \& Manufacturing}\ }\textbf {\bibinfo {volume}
{12}},\ \bibinfo {pages} {72} (\bibinfo {year} {2021})}\BibitemShut {NoStop}%
\bibitem [{\citenamefont {Dai}\ \emph {et~al.}(2020)\citenamefont {Dai},
\citenamefont {Deng}, \citenamefont {Van~Brackle}, \citenamefont {Chen},
\citenamefont {Rudd}, \citenamefont {Xiao}, \citenamefont {Lin},
\citenamefont {Chen},\ and\ \citenamefont {Huang}}]{dai2020scalable}%
\BibitemOpen
\bibfield  {author} {\bibinfo {author} {\bibfnamefont {X.}~\bibnamefont
{Dai}}, \bibinfo {author} {\bibfnamefont {Y.}~\bibnamefont {Deng}}, \bibinfo
{author} {\bibfnamefont {C.~H.}\ \bibnamefont {Van~Brackle}}, \bibinfo
{author} {\bibfnamefont {S.}~\bibnamefont {Chen}}, \bibinfo {author}
{\bibfnamefont {P.~N.}\ \bibnamefont {Rudd}}, \bibinfo {author}
{\bibfnamefont {X.}~\bibnamefont {Xiao}}, \bibinfo {author} {\bibfnamefont
{Y.}~\bibnamefont {Lin}}, \bibinfo {author} {\bibfnamefont {B.}~\bibnamefont
{Chen}},\ and\ \bibinfo {author} {\bibfnamefont {J.}~\bibnamefont {Huang}},\
}\bibfield  {title} {\bibinfo {title} {Scalable fabrication of efficient
perovskite solar modules on flexible glass substrates},\ }\href
{https://doi.org/10.1002/aenm.201903108} {\bibfield  {journal} {\bibinfo
{journal} {Advanced Energy Materials}\ }\textbf {\bibinfo {volume} {10}},\
\bibinfo {pages} {1903108} (\bibinfo {year} {2020})}\BibitemShut {NoStop}%
\bibitem [{\citenamefont {Mathur}\ \emph {et~al.}(2021)\citenamefont {Mathur},
\citenamefont {Fan},\ and\ \citenamefont
{Maheshwari}}]{mathur2021organolead}%
\BibitemOpen
\bibfield  {author} {\bibinfo {author} {\bibfnamefont {A.}~\bibnamefont
{Mathur}}, \bibinfo {author} {\bibfnamefont {H.}~\bibnamefont {Fan}},\ and\
\bibinfo {author} {\bibfnamefont {V.}~\bibnamefont {Maheshwari}},\ }\bibfield
{title} {\bibinfo {title} {Organolead halide perovskites beyond solar cells:
self-powered devices and the associated progress and challenges},\ }\href
{https://doi.org/10.1039/D1MA00377A} {\bibfield  {journal} {\bibinfo
{journal} {Materials Advances}\ }\textbf {\bibinfo {volume} {2}},\ \bibinfo
{pages} {5274} (\bibinfo {year} {2021})}\BibitemShut {NoStop}%
\bibitem [{\citenamefont {Ning}\ and\ \citenamefont
{Gao}(2019)}]{ning2019structural}%
\BibitemOpen
\bibfield  {author} {\bibinfo {author} {\bibfnamefont {W.}~\bibnamefont
{Ning}}\ and\ \bibinfo {author} {\bibfnamefont {F.}~\bibnamefont {Gao}},\
}\bibfield  {title} {\bibinfo {title} {Structural and functional diversity in
lead-free halide perovskite materials},\ }\href
{https://doi.org/10.1002/adma.201900326} {\bibfield  {journal} {\bibinfo
{journal} {Advanced Materials}\ }\textbf {\bibinfo {volume} {31}},\ \bibinfo
{pages} {1900326} (\bibinfo {year} {2019})}\BibitemShut {NoStop}%
\bibitem [{\citenamefont {Chen}\ \emph {et~al.}(2020)\citenamefont {Chen},
\citenamefont {Dong}, \citenamefont {Eickemeyer}, \citenamefont {Liu},
\citenamefont {Dai}, \citenamefont {Carl}, \citenamefont {Bahrami},
\citenamefont {Chowdhury}, \citenamefont {Grimm}, \citenamefont {Shi} \emph
{et~al.}}]{chen2020high}%
\BibitemOpen
\bibfield  {author} {\bibinfo {author} {\bibfnamefont {M.}~\bibnamefont
{Chen}}, \bibinfo {author} {\bibfnamefont {Q.}~\bibnamefont {Dong}}, \bibinfo
{author} {\bibfnamefont {F.~T.}\ \bibnamefont {Eickemeyer}}, \bibinfo
{author} {\bibfnamefont {Y.}~\bibnamefont {Liu}}, \bibinfo {author}
{\bibfnamefont {Z.}~\bibnamefont {Dai}}, \bibinfo {author} {\bibfnamefont
{A.~D.}\ \bibnamefont {Carl}}, \bibinfo {author} {\bibfnamefont
{B.}~\bibnamefont {Bahrami}}, \bibinfo {author} {\bibfnamefont {A.~H.}\
\bibnamefont {Chowdhury}}, \bibinfo {author} {\bibfnamefont {R.~L.}\
\bibnamefont {Grimm}}, \bibinfo {author} {\bibfnamefont {Y.}~\bibnamefont
{Shi}}, \emph {et~al.},\ }\bibfield  {title} {\bibinfo {title}
{High-performance lead-free solar cells based on tin-halide perovskite thin
films functionalized by a divalent organic cation},\ }\href
{https://doi.org/10.1021/acsenergylett.0c00888} {\bibfield  {journal}
{\bibinfo  {journal} {ACS Energy Letters}\ }\textbf {\bibinfo {volume} {5}},\
\bibinfo {pages} {2223} (\bibinfo {year} {2020})}\BibitemShut {NoStop}%
\bibitem [{\citenamefont {Jin}\ \emph {et~al.}(2020)\citenamefont {Jin},
\citenamefont {Zhang}, \citenamefont {Xiu}, \citenamefont {Song},
\citenamefont {Gatti},\ and\ \citenamefont {He}}]{jin2020critical}%
\BibitemOpen
\bibfield  {author} {\bibinfo {author} {\bibfnamefont {Z.}~\bibnamefont
{Jin}}, \bibinfo {author} {\bibfnamefont {Z.}~\bibnamefont {Zhang}}, \bibinfo
{author} {\bibfnamefont {J.}~\bibnamefont {Xiu}}, \bibinfo {author}
{\bibfnamefont {H.}~\bibnamefont {Song}}, \bibinfo {author} {\bibfnamefont
{T.}~\bibnamefont {Gatti}},\ and\ \bibinfo {author} {\bibfnamefont
{Z.}~\bibnamefont {He}},\ }\bibfield  {title} {\bibinfo {title} {A critical
review on bismuth and antimony halide based perovskites and their derivatives
for photovoltaic applications: recent advances and challenges},\ }\href
{https://doi.org/10.1039/D0TA05433J} {\bibfield  {journal} {\bibinfo
{journal} {Journal of Materials Chemistry A}\ }\textbf {\bibinfo {volume}
{8}},\ \bibinfo {pages} {16166} (\bibinfo {year} {2020})}\BibitemShut
{NoStop}%
\bibitem [{\citenamefont {Hoefler}\ \emph {et~al.}(2017)\citenamefont
{Hoefler}, \citenamefont {Trimmel},\ and\ \citenamefont
{Rath}}]{hoefler2017progress}%
\BibitemOpen
\bibfield  {author} {\bibinfo {author} {\bibfnamefont {S.~F.}\ \bibnamefont
{Hoefler}}, \bibinfo {author} {\bibfnamefont {G.}~\bibnamefont {Trimmel}},\
and\ \bibinfo {author} {\bibfnamefont {T.}~\bibnamefont {Rath}},\ }\bibfield
{title} {\bibinfo {title} {Progress on lead-free metal halide perovskites for
photovoltaic applications: a review},\ }\href
{https://doi.org/10.1007/s00706-017-1933-9} {\bibfield  {journal} {\bibinfo
{journal} {Monatshefte f{\"u}r Chemie-Chemical Monthly}\ }\textbf {\bibinfo
{volume} {148}},\ \bibinfo {pages} {795} (\bibinfo {year}
{2017})}\BibitemShut {NoStop}%
\bibitem [{\citenamefont {Ma}\ \emph {et~al.}(2020)\citenamefont {Ma},
\citenamefont {Wang}, \citenamefont {Ji}, \citenamefont {Chen},\ and\
\citenamefont {Shi}}]{ma2020lead}%
\BibitemOpen
\bibfield  {author} {\bibinfo {author} {\bibfnamefont {Z.}~\bibnamefont
{Ma}}, \bibinfo {author} {\bibfnamefont {L.}~\bibnamefont {Wang}}, \bibinfo
{author} {\bibfnamefont {X.}~\bibnamefont {Ji}}, \bibinfo {author}
{\bibfnamefont {X.}~\bibnamefont {Chen}},\ and\ \bibinfo {author}
{\bibfnamefont {Z.}~\bibnamefont {Shi}},\ }\bibfield  {title} {\bibinfo
{title} {Lead-free metal halide perovskites and perovskite derivatives as an
environmentally friendly emitter for light-emitting device applications},\
}\href {https://doi.org/10.1021/acs.jpclett.0c01378} {\bibfield  {journal}
{\bibinfo  {journal} {The Journal of Physical Chemistry Letters}\ }\textbf
{\bibinfo {volume} {11}},\ \bibinfo {pages} {5517} (\bibinfo {year}
{2020})}\BibitemShut {NoStop}%
\bibitem [{\citenamefont {Pecunia}\ \emph {et~al.}(2020)\citenamefont
{Pecunia}, \citenamefont {Occhipinti}, \citenamefont {Chakraborty},
\citenamefont {Pan},\ and\ \citenamefont {Peng}}]{pecunia2020lead}%
\BibitemOpen
\bibfield  {author} {\bibinfo {author} {\bibfnamefont {V.}~\bibnamefont
{Pecunia}}, \bibinfo {author} {\bibfnamefont {L.~G.}\ \bibnamefont
{Occhipinti}}, \bibinfo {author} {\bibfnamefont {A.}~\bibnamefont
{Chakraborty}}, \bibinfo {author} {\bibfnamefont {Y.}~\bibnamefont {Pan}},\
and\ \bibinfo {author} {\bibfnamefont {Y.}~\bibnamefont {Peng}},\ }\bibfield
{title} {\bibinfo {title} {Lead-free halide perovskite photovoltaics:
Challenges, open questions, and opportunities},\ }\href
{https://doi.org/10.1063/5.0022271} {\bibfield  {journal} {\bibinfo
{journal} {APL Materials}\ }\textbf {\bibinfo {volume} {8}},\ \bibinfo
{pages} {100901} (\bibinfo {year} {2020})}\BibitemShut {NoStop}%
\bibitem [{\citenamefont {Jain}\ \emph {et~al.}(2013)\citenamefont {Jain},
\citenamefont {Ong}, \citenamefont {Hautier}, \citenamefont {Chen},
\citenamefont {Richards}, \citenamefont {Dacek}, \citenamefont {Cholia},
\citenamefont {Gunter}, \citenamefont {Skinner}, \citenamefont {Ceder} \emph
{et~al.}}]{jain2013commentary}%
\BibitemOpen
\bibfield  {author} {\bibinfo {author} {\bibfnamefont {A.}~\bibnamefont
{Jain}}, \bibinfo {author} {\bibfnamefont {S.~P.}\ \bibnamefont {Ong}},
\bibinfo {author} {\bibfnamefont {G.}~\bibnamefont {Hautier}}, \bibinfo
{author} {\bibfnamefont {W.}~\bibnamefont {Chen}}, \bibinfo {author}
{\bibfnamefont {W.~D.}\ \bibnamefont {Richards}}, \bibinfo {author}
{\bibfnamefont {S.}~\bibnamefont {Dacek}}, \bibinfo {author} {\bibfnamefont
{S.}~\bibnamefont {Cholia}}, \bibinfo {author} {\bibfnamefont
{D.}~\bibnamefont {Gunter}}, \bibinfo {author} {\bibfnamefont
{D.}~\bibnamefont {Skinner}}, \bibinfo {author} {\bibfnamefont
{G.}~\bibnamefont {Ceder}}, \emph {et~al.},\ }\bibfield  {title} {\bibinfo
{title} {Commentary: The materials project: A materials genome approach to
accelerating materials innovation},\ }\href
{https://doi.org/10.1063/1.4812323} {\bibfield  {journal} {\bibinfo
{journal} {APL materials}\ }\textbf {\bibinfo {volume} {1}},\ \bibinfo
{pages} {011002} (\bibinfo {year} {2013})}\BibitemShut {NoStop}%
\bibitem [{\citenamefont {Cui}\ \emph {et~al.}(2020)\citenamefont {Cui},
\citenamefont {Li},\ and\ \citenamefont {Hu}}]{cui2020emerging}%
\BibitemOpen
\bibfield  {author} {\bibinfo {author} {\bibfnamefont {Y.}~\bibnamefont
{Cui}}, \bibinfo {author} {\bibfnamefont {M.}~\bibnamefont {Li}},\ and\
\bibinfo {author} {\bibfnamefont {Y.}~\bibnamefont {Hu}},\ }\bibfield
{title} {\bibinfo {title} {Emerging interface materials for electronics
thermal management: experiments, modeling, and new opportunities},\ }\href
{https://doi.org/10.1039/C9TC05415D} {\bibfield  {journal} {\bibinfo
{journal} {Journal of Materials Chemistry C}\ }\textbf {\bibinfo {volume}
{8}},\ \bibinfo {pages} {10568} (\bibinfo {year} {2020})}\BibitemShut
{NoStop}%
\bibitem [{\citenamefont {Xie}\ and\ \citenamefont {Grossman}(2018)}]{CGCNN}%
\BibitemOpen
\bibfield  {author} {\bibinfo {author} {\bibfnamefont {T.}~\bibnamefont
{Xie}}\ and\ \bibinfo {author} {\bibfnamefont {J.~C.}\ \bibnamefont
{Grossman}},\ }\bibfield  {title} {\bibinfo {title} {Crystal graph
convolutional neural networks for an accurate and interpretable prediction of
material properties},\ }\href
{https://doi.org/10.1103/PhysRevLett.120.145301} {\bibfield  {journal}
{\bibinfo  {journal} {Phys. Rev. Lett.}\ }\textbf {\bibinfo {volume} {120}},\
\bibinfo {pages} {145301} (\bibinfo {year} {2018})}\BibitemShut {NoStop}%
\bibitem [{\citenamefont {Kresse}\ and\ \citenamefont
{Furthm{\"u}ller}(1996)}]{kresse1996efficiency}%
\BibitemOpen
\bibfield  {author} {\bibinfo {author} {\bibfnamefont {G.}~\bibnamefont
{Kresse}}\ and\ \bibinfo {author} {\bibfnamefont {J.}~\bibnamefont
{Furthm{\"u}ller}},\ }\bibfield  {title} {\bibinfo {title} {Efficiency of
ab-initio total energy calculations for metals and semiconductors using a
plane-wave basis set},\ }\href {https://doi.org/10.1016/0927-0256(96)00008-0}
{\bibfield  {journal} {\bibinfo  {journal} {Computational materials science}\
}\textbf {\bibinfo {volume} {6}},\ \bibinfo {pages} {15} (\bibinfo {year}
{1996})}\BibitemShut {NoStop}%
\bibitem [{\citenamefont {Kresse}\ and\ \citenamefont
{Furthm\"uller}(1996)}]{kresse1996efficient}%
\BibitemOpen
\bibfield  {author} {\bibinfo {author} {\bibfnamefont {G.}~\bibnamefont
{Kresse}}\ and\ \bibinfo {author} {\bibfnamefont {J.}~\bibnamefont
{Furthm\"uller}},\ }\bibfield  {title} {\bibinfo {title} {Efficient iterative
schemes for ab initio total-energy calculations using a plane-wave basis
set},\ }\href {https://doi.org/10.1103/PhysRevB.54.11169} {\bibfield
{journal} {\bibinfo  {journal} {Phys. Rev. B}\ }\textbf {\bibinfo {volume}
{54}},\ \bibinfo {pages} {11169} (\bibinfo {year} {1996})}\BibitemShut
{NoStop}%
\bibitem [{\citenamefont {Perdew}\ \emph {et~al.}(1996)\citenamefont {Perdew},
\citenamefont {Burke},\ and\ \citenamefont {Ernzerhof}}]{PBE}%
\BibitemOpen
\bibfield  {author} {\bibinfo {author} {\bibfnamefont {J.~P.}\ \bibnamefont
{Perdew}}, \bibinfo {author} {\bibfnamefont {K.}~\bibnamefont {Burke}},\ and\
\bibinfo {author} {\bibfnamefont {M.}~\bibnamefont {Ernzerhof}},\ }\bibfield
{title} {\bibinfo {title} {Generalized gradient approximation made simple},\
}\href {https://doi.org/10.1103/PhysRevLett.77.3865} {\bibfield  {journal}
{\bibinfo  {journal} {Phys. Rev. Lett.}\ }\textbf {\bibinfo {volume} {77}},\
\bibinfo {pages} {3865} (\bibinfo {year} {1996})}\BibitemShut {NoStop}%
\bibitem [{\citenamefont {Denault}\ \emph {et~al.}(2014)\citenamefont
{Denault}, \citenamefont {Brgoch}, \citenamefont {Gaultois}, \citenamefont
{Mikhailovsky}, \citenamefont {Petry}, \citenamefont {Winkler}, \citenamefont
{DenBaars},\ and\ \citenamefont {Seshadri}}]{denault2014consequences}%
\BibitemOpen
\bibfield  {author} {\bibinfo {author} {\bibfnamefont {K.~A.}\ \bibnamefont
{Denault}}, \bibinfo {author} {\bibfnamefont {J.}~\bibnamefont {Brgoch}},
\bibinfo {author} {\bibfnamefont {M.~W.}\ \bibnamefont {Gaultois}}, \bibinfo
{author} {\bibfnamefont {A.}~\bibnamefont {Mikhailovsky}}, \bibinfo {author}
{\bibfnamefont {R.}~\bibnamefont {Petry}}, \bibinfo {author} {\bibfnamefont
{H.}~\bibnamefont {Winkler}}, \bibinfo {author} {\bibfnamefont {S.~P.}\
\bibnamefont {DenBaars}},\ and\ \bibinfo {author} {\bibfnamefont
{R.}~\bibnamefont {Seshadri}},\ }\bibfield  {title} {\bibinfo {title}
{Consequences of optimal bond valence on structural rigidity and improved
luminescence properties in {Sr$_x$Ba$_{2–x}$SiO$_4$:Eu$^{2+}$}
orthosilicate phosphors},\ }\href {https://doi.org/10.1021/cm500116u}
{\bibfield  {journal} {\bibinfo  {journal} {Chemistry of Materials}\ }\textbf
{\bibinfo {volume} {26}},\ \bibinfo {pages} {2275} (\bibinfo {year}
{2014})}\BibitemShut {NoStop}%
\bibitem [{\citenamefont {Zhuo}\ \emph {et~al.}(2018)\citenamefont {Zhuo},
\citenamefont {Mansouri~Tehrani}, \citenamefont {Oliynyk}, \citenamefont
{Duke},\ and\ \citenamefont {Brgoch}}]{zhuo2018identifying}%
\BibitemOpen
\bibfield  {author} {\bibinfo {author} {\bibfnamefont {Y.}~\bibnamefont
{Zhuo}}, \bibinfo {author} {\bibfnamefont {A.}~\bibnamefont
{Mansouri~Tehrani}}, \bibinfo {author} {\bibfnamefont {A.~O.}\ \bibnamefont
{Oliynyk}}, \bibinfo {author} {\bibfnamefont {A.~C.}\ \bibnamefont {Duke}},\
and\ \bibinfo {author} {\bibfnamefont {J.}~\bibnamefont {Brgoch}},\
}\bibfield  {title} {\bibinfo {title} {Identifying an efficient, thermally
robust inorganic phosphor host via machine learning},\ }\href
{https://doi.org/10.1038/s41467-018-06625-z} {\bibfield  {journal} {\bibinfo
{journal} {Nature communications}\ }\textbf {\bibinfo {volume} {9}},\
\bibinfo {pages} {1} (\bibinfo {year} {2018})}\BibitemShut {NoStop}%
\bibitem [{\citenamefont {Ullrich}\ \emph {et~al.}(2017)\citenamefont
{Ullrich}, \citenamefont {Bhowmick},\ and\ \citenamefont
{Xi}}]{ullrich2017relation}%
\BibitemOpen
\bibfield  {author} {\bibinfo {author} {\bibfnamefont {B.}~\bibnamefont
{Ullrich}}, \bibinfo {author} {\bibfnamefont {M.}~\bibnamefont {Bhowmick}},\
and\ \bibinfo {author} {\bibfnamefont {H.}~\bibnamefont {Xi}},\ }\bibfield
{title} {\bibinfo {title} {Relation between debye temperature and energy band
gap of semiconductors},\ }\href {https://doi.org/10.1063/1.4980142}
{\bibfield  {journal} {\bibinfo  {journal} {AIP Advances}\ }\textbf {\bibinfo
{volume} {7}},\ \bibinfo {pages} {045109} (\bibinfo {year}
{2017})}\BibitemShut {NoStop}%
\bibitem [{\citenamefont {Hautier}\ \emph {et~al.}(2011)\citenamefont
{Hautier}, \citenamefont {Fischer}, \citenamefont {Ehrlacher}, \citenamefont
{Jain},\ and\ \citenamefont {Ceder}}]{hautier2011data}%
\BibitemOpen
\bibfield  {author} {\bibinfo {author} {\bibfnamefont {G.}~\bibnamefont
{Hautier}}, \bibinfo {author} {\bibfnamefont {C.}~\bibnamefont {Fischer}},
\bibinfo {author} {\bibfnamefont {V.}~\bibnamefont {Ehrlacher}}, \bibinfo
{author} {\bibfnamefont {A.}~\bibnamefont {Jain}},\ and\ \bibinfo {author}
{\bibfnamefont {G.}~\bibnamefont {Ceder}},\ }\bibfield  {title} {\bibinfo
{title} {Data mined ionic substitutions for the discovery of new compounds},\
}\href {https://doi.org/10.1021/ic102031h} {\bibfield  {journal} {\bibinfo
{journal} {Inorganic chemistry}\ }\textbf {\bibinfo {volume} {50}},\ \bibinfo
{pages} {656} (\bibinfo {year} {2011})}\BibitemShut {NoStop}%
\bibitem [{\citenamefont {Kim}\ \emph {et~al.}(2023)\citenamefont {Kim},
\citenamefont {Han}, \citenamefont {Ko}, \citenamefont {Samanta},
\citenamefont {Lee}, \citenamefont {Jeon}, \citenamefont {Kim}, \citenamefont
{Chung}, \citenamefont {Im},\ and\ \citenamefont {Cho}}]{kim2023high}%
\BibitemOpen
\bibfield  {author} {\bibinfo {author} {\bibfnamefont {H.~W.}\ \bibnamefont
{Kim}}, \bibinfo {author} {\bibfnamefont {J.~H.}\ \bibnamefont {Han}},
\bibinfo {author} {\bibfnamefont {H.}~\bibnamefont {Ko}}, \bibinfo {author}
{\bibfnamefont {T.}~\bibnamefont {Samanta}}, \bibinfo {author} {\bibfnamefont
{D.~G.}\ \bibnamefont {Lee}}, \bibinfo {author} {\bibfnamefont {D.~W.}\
\bibnamefont {Jeon}}, \bibinfo {author} {\bibfnamefont {W.}~\bibnamefont
{Kim}}, \bibinfo {author} {\bibfnamefont {Y.-C.}\ \bibnamefont {Chung}},
\bibinfo {author} {\bibfnamefont {W.~B.}\ \bibnamefont {Im}},\ and\ \bibinfo
{author} {\bibfnamefont {S.~B.}\ \bibnamefont {Cho}},\ }\bibfield  {title}
{\bibinfo {title} {High-throughput screening on halide perovskite derivatives
and rational design of {Cs$_3$LuCl$_6$}},\ }\href
{https://doi.org/10.1021/acsenergylett.3c01207} {\bibfield  {journal}
{\bibinfo  {journal} {ACS Energy Letters}\ }\textbf {\bibinfo {volume} {8}},\
\bibinfo {pages} {3621} (\bibinfo {year} {2023})}\BibitemShut {NoStop}%
\bibitem [{\citenamefont {Kumar}\ \emph {et~al.}(2024)\citenamefont {Kumar},
\citenamefont {Kim}, \citenamefont {Singh}, \citenamefont {Cho},\ and\
\citenamefont {Ko}}]{kumar2024designing}%
\BibitemOpen
\bibfield  {author} {\bibinfo {author} {\bibfnamefont {U.}~\bibnamefont
{Kumar}}, \bibinfo {author} {\bibfnamefont {H.~W.}\ \bibnamefont {Kim}},
\bibinfo {author} {\bibfnamefont {S.}~\bibnamefont {Singh}}, \bibinfo
{author} {\bibfnamefont {S.~B.}\ \bibnamefont {Cho}},\ and\ \bibinfo {author}
{\bibfnamefont {H.}~\bibnamefont {Ko}},\ }\bibfield  {title} {\bibinfo
{title} {Designing pr-based advanced photoluminescent materials using machine
learning and density functional theory},\ }\href
{https://doi.org/10.1007/s10853-023-09232-6} {\bibfield  {journal} {\bibinfo
{journal} {Journal of Materials Science}\ ,\ \bibinfo {pages} {1}} (\bibinfo
{year} {2024})}\BibitemShut {NoStop}%
\bibitem [{\citenamefont {Peterson}\ and\ \citenamefont
{Brgoch}(2021)}]{peterson2021materials}%
\BibitemOpen
\bibfield  {author} {\bibinfo {author} {\bibfnamefont {G.~G.}\ \bibnamefont
{Peterson}}\ and\ \bibinfo {author} {\bibfnamefont {J.}~\bibnamefont
{Brgoch}},\ }\bibfield  {title} {\bibinfo {title} {Materials discovery
through machine learning formation energy},\ }\href
{https://doi.org/10.1088/2515-7655/abe425} {\bibfield  {journal} {\bibinfo
{journal} {Journal of Physics: Energy}\ }\textbf {\bibinfo {volume} {3}},\
\bibinfo {pages} {022002} (\bibinfo {year} {2021})}\BibitemShut {NoStop}%
\bibitem [{\citenamefont {Shockley}\ and\ \citenamefont
{Queisser}(2018)}]{shockley2018detailed}%
\BibitemOpen
\bibfield  {author} {\bibinfo {author} {\bibfnamefont {W.}~\bibnamefont
{Shockley}}\ and\ \bibinfo {author} {\bibfnamefont {H.}~\bibnamefont
{Queisser}},\ }\bibfield  {title} {\bibinfo {title} {Detailed balance limit
of efficiency of p--n junction solar cells},\ }in\ \href@noop {} {\emph
{\bibinfo {booktitle} {Renewable Energy}}}\ (\bibinfo  {publisher}
{Routledge},\ \bibinfo {year} {2018})\ pp.\ \bibinfo {pages}
{Vol2\_35--Vol2\_54}\BibitemShut {NoStop}%
\bibitem [{\citenamefont {Ling}\ \emph {et~al.}(2022)\citenamefont {Ling},
\citenamefont {Montoya}, \citenamefont {Hung},\ and\ \citenamefont
{Aykol}}]{ling2022solving}%
\BibitemOpen
\bibfield  {author} {\bibinfo {author} {\bibfnamefont {H.}~\bibnamefont
{Ling}}, \bibinfo {author} {\bibfnamefont {J.}~\bibnamefont {Montoya}},
\bibinfo {author} {\bibfnamefont {L.}~\bibnamefont {Hung}},\ and\ \bibinfo
{author} {\bibfnamefont {M.}~\bibnamefont {Aykol}},\ }\bibfield  {title}
{\bibinfo {title} {Solving inorganic crystal structures from x-ray powder
diffraction using a generative first-principles framework},\ }\href
{https://doi.org/10.1016/j.commatsci.2022.111687} {\bibfield  {journal}
{\bibinfo  {journal} {Computational Materials Science}\ }\textbf {\bibinfo
{volume} {214}},\ \bibinfo {pages} {111687} (\bibinfo {year}
{2022})}\BibitemShut {NoStop}%
\bibitem [{\citenamefont {Toll}(1956)}]{toll1956causality}%
\BibitemOpen
\bibfield  {author} {\bibinfo {author} {\bibfnamefont {J.~S.}\ \bibnamefont
{Toll}},\ }\bibfield  {title} {\bibinfo {title} {Causality and the dispersion
relation: logical foundations},\ }\href
{https://doi.org/10.1103/PhysRev.104.1760} {\bibfield  {journal} {\bibinfo
{journal} {Physical review}\ }\textbf {\bibinfo {volume} {104}},\ \bibinfo
{pages} {1760} (\bibinfo {year} {1956})}\BibitemShut {NoStop}%
\bibitem [{\citenamefont {Ehrenreich}\ and\ \citenamefont
{Cohen}(1959)}]{ehrenreich1959self}%
\BibitemOpen
\bibfield  {author} {\bibinfo {author} {\bibfnamefont {H.}~\bibnamefont
{Ehrenreich}}\ and\ \bibinfo {author} {\bibfnamefont {M.~H.}\ \bibnamefont
{Cohen}},\ }\bibfield  {title} {\bibinfo {title} {Self-consistent field
approach to the many-electron problem},\ }\href
{https://doi.org/10.1103/PhysRev.115.786} {\bibfield  {journal} {\bibinfo
{journal} {Physical Review}\ }\textbf {\bibinfo {volume} {115}},\ \bibinfo
{pages} {786} (\bibinfo {year} {1959})}\BibitemShut {NoStop}%
\bibitem [{\citenamefont {Wang}\ \emph {et~al.}(2014)\citenamefont {Wang},
\citenamefont {Xiao}, \citenamefont {Ma}, \citenamefont {Liu},\ and\
\citenamefont {Yang}}]{wang2014structural}%
\BibitemOpen
\bibfield  {author} {\bibinfo {author} {\bibfnamefont {V.}~\bibnamefont
{Wang}}, \bibinfo {author} {\bibfnamefont {W.}~\bibnamefont {Xiao}}, \bibinfo
{author} {\bibfnamefont {D.-M.}\ \bibnamefont {Ma}}, \bibinfo {author}
{\bibfnamefont {R.-J.}\ \bibnamefont {Liu}},\ and\ \bibinfo {author}
{\bibfnamefont {C.-M.}\ \bibnamefont {Yang}},\ }\bibfield  {title} {\bibinfo
{title} {Structural, electronic, and optical properties of gaino3: A hybrid
density functional study},\ }\href {https://doi.org/10.1063/1.4863210}
{\bibfield  {journal} {\bibinfo  {journal} {Journal of Applied Physics}\
}\textbf {\bibinfo {volume} {115}},\ \bibinfo {pages} {043708} (\bibinfo
{year} {2014})}\BibitemShut {NoStop}%
\bibitem [{\citenamefont {Huang}\ \emph {et~al.}(2014)\citenamefont {Huang},
\citenamefont {Fang}, \citenamefont {Omenya}, \citenamefont {O'Shea},
\citenamefont {Chernova}, \citenamefont {Zhang}, \citenamefont {Wang},
\citenamefont {Quackenbush}, \citenamefont {Piper}, \citenamefont {Scanlon}
\emph {et~al.}}]{huang2014understanding}%
\BibitemOpen
\bibfield  {author} {\bibinfo {author} {\bibfnamefont {Y.}~\bibnamefont
{Huang}}, \bibinfo {author} {\bibfnamefont {J.}~\bibnamefont {Fang}},
\bibinfo {author} {\bibfnamefont {F.}~\bibnamefont {Omenya}}, \bibinfo
{author} {\bibfnamefont {M.}~\bibnamefont {O'Shea}}, \bibinfo {author}
{\bibfnamefont {N.~A.}\ \bibnamefont {Chernova}}, \bibinfo {author}
{\bibfnamefont {R.}~\bibnamefont {Zhang}}, \bibinfo {author} {\bibfnamefont
{Q.}~\bibnamefont {Wang}}, \bibinfo {author} {\bibfnamefont {N.~F.}\
\bibnamefont {Quackenbush}}, \bibinfo {author} {\bibfnamefont {L.~F.}\
\bibnamefont {Piper}}, \bibinfo {author} {\bibfnamefont {D.~O.}\ \bibnamefont
{Scanlon}}, \emph {et~al.},\ }\bibfield  {title} {\bibinfo {title}
{Understanding the stability of mnpo 4},\ }\href
{https://doi.org/10.1039/C4TA00434E} {\bibfield  {journal} {\bibinfo
{journal} {Journal of Materials Chemistry A}\ }\textbf {\bibinfo {volume}
{2}},\ \bibinfo {pages} {12827} (\bibinfo {year} {2014})}\BibitemShut
{NoStop}%
\bibitem [{\citenamefont {Feng}\ \emph {et~al.}(2023)\citenamefont {Feng},
\citenamefont {Wang}, \citenamefont {Liu},\ and\ \citenamefont
{Yang}}]{feng2023first}%
\BibitemOpen
\bibfield  {author} {\bibinfo {author} {\bibfnamefont {Y.}~\bibnamefont
{Feng}}, \bibinfo {author} {\bibfnamefont {Z.}~\bibnamefont {Wang}}, \bibinfo
{author} {\bibfnamefont {N.}~\bibnamefont {Liu}},\ and\ \bibinfo {author}
{\bibfnamefont {Q.}~\bibnamefont {Yang}},\ }\bibfield  {title} {\bibinfo
{title} {First-principles prediction of two-dimensional mnox (x= cl, br)
monolayers: the half-metallic multiferroics with magnetoelastic coupling},\
}\href {https://doi.org/10.1039/D2NR05764F} {\bibfield  {journal} {\bibinfo
{journal} {Nanoscale}\ }\textbf {\bibinfo {volume} {15}},\ \bibinfo {pages}
{4546} (\bibinfo {year} {2023})}\BibitemShut {NoStop}%
\bibitem [{\citenamefont {Chen}\ \emph {et~al.}(2013)\citenamefont {Chen},
\citenamefont {Liu}, \citenamefont {Xiang}, \citenamefont {Feng},\ and\
\citenamefont {Qiu}}]{chen2013synthesis}%
\BibitemOpen
\bibfield  {author} {\bibinfo {author} {\bibfnamefont {S.}~\bibnamefont
{Chen}}, \bibinfo {author} {\bibfnamefont {F.}~\bibnamefont {Liu}}, \bibinfo
{author} {\bibfnamefont {Q.}~\bibnamefont {Xiang}}, \bibinfo {author}
{\bibfnamefont {X.}~\bibnamefont {Feng}},\ and\ \bibinfo {author}
{\bibfnamefont {G.}~\bibnamefont {Qiu}},\ }\bibfield  {title} {\bibinfo
{title} {Synthesis of {Mn$_2$O$_3$} microstructures and their energy storage
ability studies},\ }\href {https://doi.org/10.1016/j.electacta.2013.06.001}
{\bibfield  {journal} {\bibinfo  {journal} {Electrochimica Acta}\ }\textbf
{\bibinfo {volume} {106}},\ \bibinfo {pages} {360} (\bibinfo {year}
{2013})}\BibitemShut {NoStop}%
\bibitem [{\citenamefont {Liu}\ \emph {et~al.}(2024)\citenamefont {Liu},
\citenamefont {Zhang}, \citenamefont {Xiong}, \citenamefont {Pang},
\citenamefont {Liu}, \citenamefont {Yang}, \citenamefont {Yu}, \citenamefont
{Li}, \citenamefont {Zhu},\ and\ \citenamefont {Wu}}]{liu2024spin}%
\BibitemOpen
\bibfield  {author} {\bibinfo {author} {\bibfnamefont {B.}~\bibnamefont
{Liu}}, \bibinfo {author} {\bibfnamefont {X.}~\bibnamefont {Zhang}}, \bibinfo
{author} {\bibfnamefont {J.}~\bibnamefont {Xiong}}, \bibinfo {author}
{\bibfnamefont {X.}~\bibnamefont {Pang}}, \bibinfo {author} {\bibfnamefont
{S.}~\bibnamefont {Liu}}, \bibinfo {author} {\bibfnamefont {Z.}~\bibnamefont
{Yang}}, \bibinfo {author} {\bibfnamefont {Q.}~\bibnamefont {Yu}}, \bibinfo
{author} {\bibfnamefont {H.}~\bibnamefont {Li}}, \bibinfo {author}
{\bibfnamefont {S.}~\bibnamefont {Zhu}},\ and\ \bibinfo {author}
{\bibfnamefont {J.}~\bibnamefont {Wu}},\ }\bibfield  {title} {\bibinfo
{title} {Spin transport of half-metal mn2x3 with high curie temperature: An
ideal giant magnetoresistance device from electrical and thermal drives},\
}\href {https://doi.org/10.1007/s11467-023-1367-2} {\bibfield  {journal}
{\bibinfo  {journal} {Frontiers of Physics}\ }\textbf {\bibinfo {volume}
{19}},\ \bibinfo {pages} {1} (\bibinfo {year} {2024})}\BibitemShut {NoStop}%
\end{thebibliography}
%

\end{document}